\documentclass[twocolumn]{aastex62}

\newcommand{\OII}{[O\tiny{ }\footnotesize{II}\normalsize{] }}

\newcommand{\MgII}{Mg\tiny{ }\footnotesize{II}\normalsize{ }}
\newcommand{\SiII}{Si\tiny{ }\footnotesize{II}\normalsize{ }}
\newcommand{\FeII}{Fe\tiny{ }\footnotesize{II}\normalsize{ }}

\newcommand{\kms}{\ifmmode\,{\rm km}\,{\rm s}^{-1}\else km$\,$s$^{-1}$\fi}
\newcommand{\Msun}{\mathrm{M}_{\sun}}
\newcommand{\sfrunit}{\mathrm{M}_{\sun}~\mathrm{yr}^{-1}}

\newcommand{\comment}[1]{\textcolor{black}{#1}}

\defcitealias{Dopita16}{D16}

\graphicspath{{./}{}}

\shorttitle{\MgII Tomography of a $z=0.77$ Galaxy Halo}
\shortauthors{Mortensen et al.}
\begin{document}

% \title{Reducing systematic uncertainties in strong-line metallicity
% measurements:  a case study in a low mass galaxy at $z\sim1$}

\title{
Kinematics of the Circumgalactic Medium\\of a $z=0.77$ Galaxy from \MgII Tomography
}

\correspondingauthor{Kris Mortensen}
\email{kristophermortensen2020@u.northwestern.edu}

\author{Kris Mortensen}
\affiliation{Department of Physics, University of California, Davis, 1 Shields Avenue, Davis, CA 95616, USA}

\author{Keerthi Vasan G.C.}
\affiliation{Department of Physics, University of California, Davis, 1 Shields Avenue, Davis, CA 95616, USA}

\author{Tucker Jones}
\affiliation{Department of Physics, University of California, Davis, 1 Shields Avenue, Davis, CA 95616, USA}

\author{Claude-Andr\'{e} Faucher-Gigu\`{e}re}
\affiliation{Department of Physics and Astronomy, Northwestern University, 2145 Sheridan Road, Evanston, IL 60201, USA}

\author{Ryan Sanders}
\affiliation{Department of Physics, University of California, Davis, 1 Shields Avenue, Davis, CA 95616, USA}

\author{Richard S. Ellis}
\affiliation{Department of Physics and Astronomy, University College London, Gower Street, London WC1E 6BT, UK}

\author{Nicha Leethochawalit}
\affiliation{School of Physics, The University of Melbourne, Parkville, VIC 3010, Australia}
\affiliation{ARC Centre of Excellence for All Sky Astrophysics in 3 Dimensions (ASTRO 3D), Australia}

\author{Daniel P. Stark}
\affiliation{Steward Observatory, University of Arizona, 933 N Cherry Ave, Tucson, AZ 85721, USA}



\begin{abstract}

Galaxy evolution is thought to be driven in large part by the flow of gas between galaxies and the circumgalactic medium (CGM), a halo of metal-enriched gas extending out to $\gtrsim100$ kpc from each galaxy. Studying the spatial structure of the CGM holds promise for understanding these gas flow mechanisms; however, the common method using background quasar sightlines provides minimal spatial information. Recent works have shown the utility of extended background sources such as giant gravitationally-lensed arcs. Using background lensed arcs from the CSWA 38 lens system, we continuously probed, at a resolution element of about 15\,kpc$^2$, the spatial and kinematic distribution of \MgII absorption in a star-forming galaxy at $z=0.77$ (stellar mass $\approx10^{9.7}~\Msun$, star formation rate~$\approx10~\sfrunit$) at impact parameters \textcolor{black}{$D \simeq 5$--30\,kpc}. Our results present an anisotropic, optically thick medium whose absorption strength decreases with increasing impact parameter, in agreement with the statistics towards quasars and other gravitational arcs. \textcolor{black}{Furthermore, we find generally low line-of-sight velocities in comparison to the relatively high velocity dispersion in the \MgII gas (with typical $\sigma\approx50~\kms$). While the galaxy itself exhibits a clear outflow (with \MgII velocities up to $\sim 500 \kms$) in the down-the-barrel spectrum, the outflow component is sub-dominant and only weakly detected at larger impact parameters probed by the background arcs. 
Our results provide evidence of mainly \textcolor{black}{dispersion-supported}, metal-enriched gas recycling through the CGM.}

\end{abstract}



\section{Introduction} \label{sec:intro}
Observing the distribution and kinematics of gas within galaxies is a major challenge in understanding galactic evolution. Much of the activity which drives a galaxy's evolution occurs in the circumgalactic medium \citep[CGM; e.g.,][]{Angles-Alcazar2017, Hafen2019a}, a region between a galaxy and the intergalactic medium \citep{Rudie2012, Shull2014}. The rate of gas accretion from the CGM is the primary driver of the star formation rate (SFR) in galaxies \citep[e.g.,][]{Dekel2009, vandeVoort2011}. Accretion of CGM gas with modest metallicity also explains the relative paucity of low metallicity stars within the disk \citep{vandenBergh1962, Sommer-Larsen1991, WolfWest2012} as well as the existence of high-column but low-metallicity IGM absorbers \citep[e.g.,][]{Lehner2013, Hafen2017}.

%% Moved figure here so that Lens model image and the table stayed in one page. - Keerthi
\begin{figure*}[ht!]

%Keerthi - Replaced the following code 
% \centerline{
% \includegraphics[trim=1cm 1cm 1cm 1cm, height=10cm]{cswa38_HST_annotated_rgb.pdf}
% \includegraphics[trim=0.25cm 0.25cm 0.25cm 0.25cm, height=8cm]{cswa38_HST_zoom.pdf}
% }
% \centerline{
% \includegraphics[width = \textwidth]{whitelightimage.png}
% }
% to:
\centerline{
\includegraphics[width =0.9\textwidth]{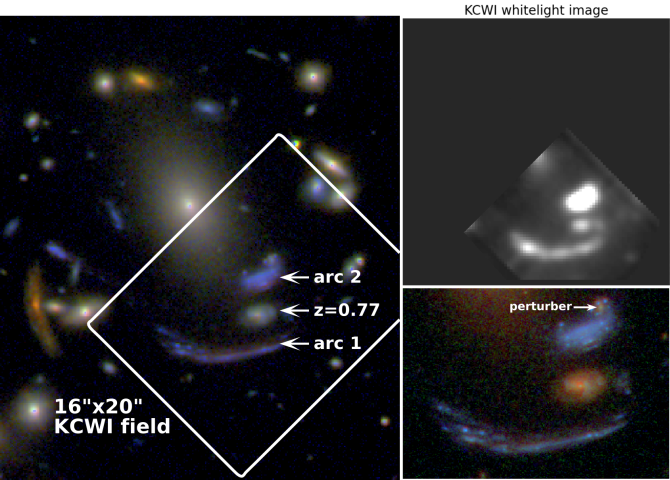}
}
\caption{
Color HST images of the CSWA 38 lens system. North is up and east is to the left. 
{\em Left:} F160W/F110W/F814W color image with the KCWI field of view shown for scale \comment{(oriented at a sky position angle of 135 degrees)}. The $z=2.92$ arcs and $z=0.77$ \MgII absorption host are labeled. \comment{{\em Top Right:} KCWI white light image centered on the absorption host and background arcs.}
{\em Bottom Right:} F110W/F814W/F606W image zoomed in on the $z=0.77$ target and background arcs. A perturber galaxy is evident in HST imaging near arc 2, creating multiple images of region C (see Figure~\ref{fig:critical_curve}), and its mass profile is included in the lens model.
} \label{fig:image}
\end{figure*}

%%%%%%%%%%%%

CGM regions typically have been probed at 10--100\,kpc scales through absorption seen in the spectra of background quasars \citep[e.g.,][]{Nielsen2013b,Prochaska2017,Tumlinson2017,Chen2017} and galaxies \citep[e.g.,][]{Steidel2010,Diamond-Stanic2016,Rubin2018b}. 
These techniques have yielded critical observational constraints and evidence towards the relationship between the CGM and galaxy properties; however, the data gathered from such probes rarely provide conclusions on the spatial structure within each CGM. Unless a galaxy has the rare privilege of multiple sight lines piercing through its CGM \citep[e.g.,][]{Lehner2020}, spatial information requires either stacking the spectra of background sources \citep[e.g.,][]{Steidel2010,Bordoloi2011,Rubin2018a,Rubin2018c} or averaging ensembles of absorber properties \citep{Chen2010,Nielsen2013a}. 
While such statistical studies provide important information regarding the average CGM profile around different galaxy populations, they provide only a crude view of the CGM structure around individual galaxies. \textcolor{black}{Some effort has been made to probe the CGM of intervening galaxies through the use of multiple sightlines, especially multiply-imaged lensed quasars \citep{Smette1992, Lopez1999, Lopez2005, Lopez2007, Rauch2001, Ellison2004, Chen2014, Zahedy2016, Rubin2018c, Zabl2020}. While these quasars help to resolve kpc scales within the CGM, the scarcity of such observed objects results in small samples and a limited sampling relative to the overall CGM areal extent.}
However, recent observations of extended gravitationally lensed arcs \citep{Lopez2018, Lopez2020} provide enhanced spatial sampling of CGM, probing gaseous halos in individual galaxies on scales of 1--100~kpc without potential biases or loss of information introduced by stacking techniques.

In this paper, we probe the the spatial and kinematic distribution of \MgII in the CGM of a $z=0.77$ galaxy based on spatially resolved spectroscopy of CSWA 38, one of the gravitational lens systems catalogued in the Cambridge
And Sloan Survey Of Wide ARcs in the skY (CASSOWARY, with target names abbreviated as CSWA; \citealt{Belokurov2007,Belokurov2009}). CSWA 38 consists of a galaxy cluster at $z=0.43$ with two luminous giant arcs at $z\approx2.92$, and multiple other moderately magnified background sources \citep[Figure~\ref{fig:image};][]{Koester2010,Bayliss2011}. The subject of this study is a $z=0.77$ galaxy which lies between the two giant arcs. Moderate resolution spectroscopy revealed prominent \MgII and \FeII absorption at $z=0.77$ in the background arc spectra \citep{Jones2018}, making this system an ideal candidate for CGM absorption tomography presented herein.

The paper is structured as follows. In Sections \ref{sec:data} and \ref{sec:lens_model}, we describe spectroscopic observations of the lensing system as well as the lens model used to de-magnify the absorber galaxy and to calculate impact parameters in the absorber plane. Sections \ref{sec:analysis} and \ref{sec:source_plane} present the main analysis of the absorber galaxy properties, and the line strength and kinematics of the \MgII gas. We discuss our results in Section \ref{sec:model} and present our summary and conclusions in Section \ref{sec:discussion}. Throughout this paper we adopt a flat $\Lambda$CDM cosmology with $H_0 = 70$~km~s$^{-1}$~Mpc$^{-1}$, $\Omega_{\rm m} = 0.26$, and $\Omega_{\Lambda} = 0.74$.
%CA: I removed '0.01' in H0 value -- not significant digits.
%\CA{are all the lengths/distances/impact parameters in the paper given in proper kpc? or are some in comoving kpc? the paper usually doesn't specify. if all in proper kpc, should state clearly here.}

\section{Spectroscopic Data} \label{sec:data}

CSWA 38 was observed with the Keck Cosmic Web Imager \citep[KCWI;][]{Morrissey2018} on \comment{four nights during three separate observing runs. Two orthogonal sky position angles (PA) were used with comparable depth in each. We observed at a PA of 135 degrees on 2018 June 17 (90 minutes on-source) and 2020 June 20 (100 minutes), and at a PA of 45 degrees on 2019 June 2 (60 minutes) and 2020 June 19 (97 minutes). The total on-source exposure time is thus 347 minutes or 5.8 hours. 
Two exposures (40 minutes on-source) taken on 2020 June 19 were offset to cover the eastern counter image of arc 2, while the remaining exposures were approximately centered on the $z=0.77$ absorber galaxy and two bright arcs (see Figures~\ref{fig:image} and \ref{fig:critical_curve}). 
}

Individual exposure times were \comment{600--1200 seconds}. Conditions ranged from clear to 0.5 magnitude of cloud extinction, with 0.8\arcsec--1.1\arcsec~seeing. KCWI was configured with the medium slicer (0.7\arcsec~slit width), BL grating, and central wavelength $\lambda_c=5150$\,\AA\ with the blue blocking filter retracted. This provides wavelength coverage from approximately 4000--6300\,\AA. This range includes the \MgII$\lambda\lambda$2796,2802 doublet redshifted to $\sim$4950\,\AA\ at $z=0.77$, and several \FeII\ transitions at rest-frame 2344--2600\,\AA. From arc lamp exposures we measure an approximately constant spectral FWHM~$=2.40$\,\AA\ with $<$\,5\% variation across the full wavelength range (e.g., $R=2060$ at $\lambda=4950$\,\AA). 

Data were reduced using the KCWI data reduction pipeline (KDERP) version 1.0.2. KDERP performs instrument signature removal (bias, dark current, scattered light, and flat fielding), sky subtraction, wavelength calibration, and spatial rectification including a correction for differential atmospheric refraction. Output data cubes have 0.68\arcsec$\times$0.29\arcsec~spatial pixels. 
Observations of the standard stars \comment{HZ43, BD25D3941, BD26d2608 and g93-48} were taken on the same nights and used for flux calibration of the \comment{2018, 2019 and 2020} data, respectively. \comment{The pipeline-reduced datacubes have a non-zero residual background with a spatial gradient, which affects measurements of absorption equivalent width if not properly corrected. We model this residual structure with a two-dimensional-first order polynomial (i.e. a plane), fit to blank sky regions in a pseudo-image generated by taking the median flux in each spaxel over the wavelength range around the \MgII absorption ($\lambda = 4834$--5084~\AA). We subtract this fit from each wavelength slice of the datacube. This approach is similar to the correction described by \cite{Burchett2020} and we find that it adequately removes the residual spatial background structure at the wavelengths of interest. 
}

\comment{ The sky-corrected datacubes from individual exposures were resampled to a common grid and combined with a weighted mean.} The resulting datacube has 0.3\arcsec~spatial pixels, adequately sampling both the native pixel size and the seeing. This final datacube is used for all subsequent analysis. 
\comment{A ``white light'' image of this datacube, created by summing all pixels in the wavelength direction, is shown in Figure~\ref{fig:image}.}

\begin{deluxetable}{lcc}[b!]
\tablecaption{RA and Dec Positions of Objects Included in the Lens Model
\label{tab:Ra-dec-information}
}
\tablecolumns{3}
% \tablenum{1}
\tablewidth{0pt}
\tablehead{
\colhead{Object} &
\colhead{$\alpha$}&
\colhead{$\delta$}
}
\startdata
cD+Cluster ($z=0.43$)  & 12:26:51.7 &  +21:52:25.4  \\
Perturber ($z=0.43$)  &  12:26:51.2574 & +21:52:21.214\\ 
Absorber ($z=0.77$)   & 12:26:51.3325  & +21:52:17.154 
\enddata
\tablecomments{Galaxy centroid positions are determined from HST optical (F606W) images. The perturber redshift is unknown and assumed to be at the same $z=0.43$ as the cluster.}
\end{deluxetable}

\section{Gravitational Lens Model} \label{sec:lens_model}

\begin{figure*}[!ph]

\centerline{
\includegraphics[width = \textwidth]{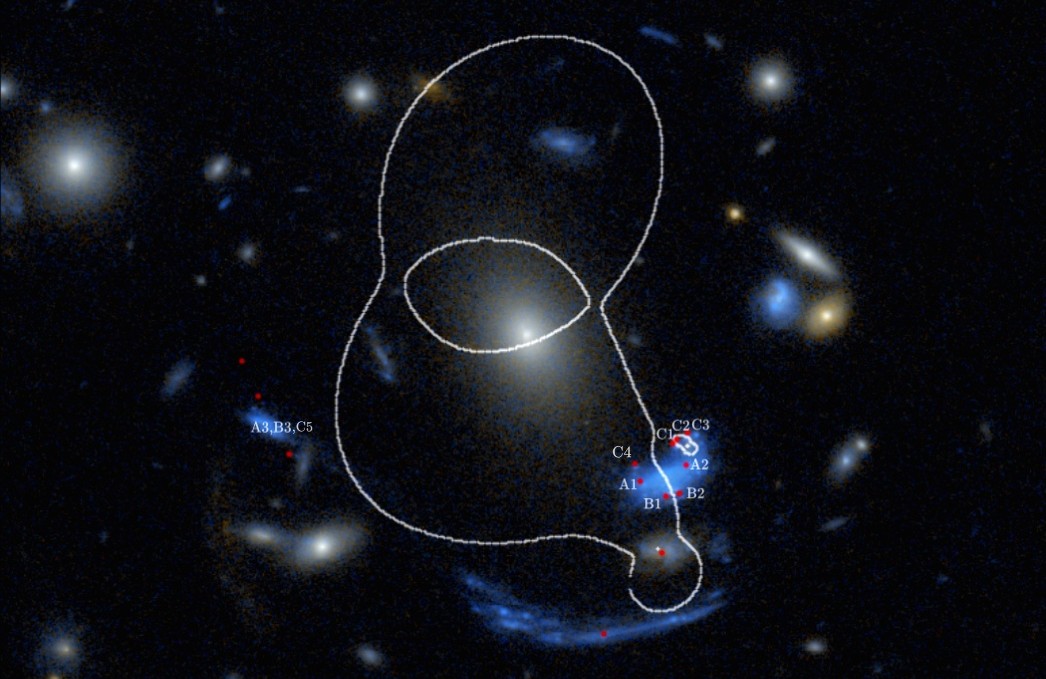}
}

\caption{
HST image with the $z=2.92$ critical curve obtained from the lens model superimposed as the white line. Multiple images of individual regions (e.g. region A: A1, A2, A3) are used as constraints in the lens model. Red points show the model-predicted positions which are accurately reproduced in the northwestern arc (A1-A2, B1-B2, C1-C2-C3-C4). The eastern counter image (A3, B3, C5) is reproduced in the correct vicinity although less accurately, likely in part because of nearby substructure which is not included in the model. 
} \label{fig:critical_curve}
\end{figure*}
\begin{table*}[!ph]

\tablecolumns{4}

\begin{tabular}{lcccccccccccccccccc}
Component & $\sigma$ & $\sigma_{\text{prior}}$ & x & x$_{prior}$ & y & y$_{prior}$ & e  & e$_{prior}$ & PA  & PA$_{prior}$ \\
and Profile & or $M$  & or $M_{\text{prior}}$   &  & '' & & '' &  &  &   &  &  & \\
\hline
cD-SIE       & 490 [\kms]  &   G(430,100)&  0.37 & G(0,0.2)  & -0.34 & G(0,0.2) & 0.41 & U(0.2,0.5) & 22 & U(10,30)  \\
Cluster-NFW  & $3.0\times10^{14}$  [$h^{-1} \Msun$]  & G(3e+14,2e+14)   & 1.25 &  U(-3,3)  &  1.94 & U(-3,3)  & 0.39 & U(0.2,0.5)  & 147  & - \\ 
Perturber-SIE  &  59 [\kms] & - &  -6.2  & -  & -4.3 & - &  0.4 & U(0.2,0.4)  & 82 & - \\ 
Absorber-SIS & 160 [\kms] &   G(80,10) &  -5.06 & -  & -8.23  & -  & - & - & - & -  
\end{tabular}

\caption{
Best-fit parameters of the lens model. $x,y$ are the coordinates of the center of each profile in arcseconds relative to the central deflector galaxy (Table~\ref{tab:Ra-dec-information}), with North up (y) and East left (x). $e$ is ellipticity and PA is the sky position angle. 
The absorber SIS $\sigma$ value scaled to the $z=0.43$ plane is $160 \kms$. 
% (corresponding to 181 $\kms$ in the $z=0.77$ plane). 
The priors used for each parameter are also listed.  A Gaussian prior with a standard deviation $\sigma$ is denoted as $G(\text{value}, \sigma)$, an uniform prior from $a$--$b$ is denoted as $U(a,b)$, and '-' indicates that no priors were used. 
\label{tab:lens model parameters}}
%The absorber SIS $\sigma$ value is scaled to the $z=0.43$ plane; {\bf the value at z=0.77 is ???. ). 
%\CA{what are the different rows in this table? different models fit to the same potential? or one potential per halo included in the overall lens model? should we specify the redshift of each halo? ... I'm unclear how to read this table.}}

\end{table*}

\begin{figure*}[htb!]

\centerline{
  \includegraphics[width=\linewidth]{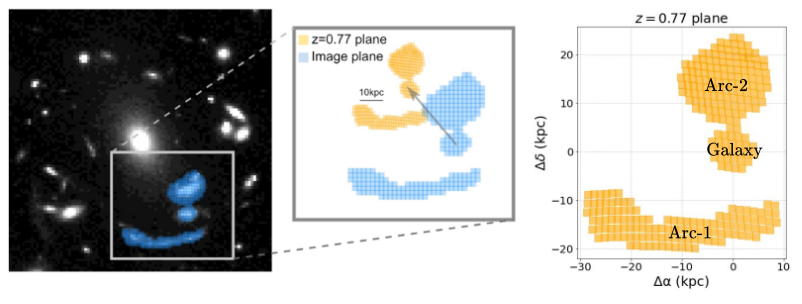}
}

\caption{
\comment{ 
 Reconstruction of the absorber galaxy and the arcs to the absorber plane ($z=0.77$) using the lens model described in Section~\ref{sec:lens_model}. The region shaded in blue indicates the spaxels in both arcs and the absober galaxy with continuum S/N $>5\sigma$; shown in orange is the same region in the $z=0.77$ plane. The gray arrow demonstrates the reconstruction of the center of the absorber galaxy from the image plane to the $z=0.77$ plane. 
 The scale bar shows 10 kpc in the $z=0.77$ plane.}
 %\TJ{(Move the individual spaxel spectra [right] to a separate figure?)}
} \label{fig:reconstruction}
\end{figure*}

In order to study the impact parameter of the CGM probed by the arcs, we must account for gravitational lensing of the region around the \MgII absorber galaxy. In this section we describe the adopted lens model which sufficiently reproduces all the observational constraints. We model the lens as a combination of a galaxy cluster-scale dark matter halo plus individual galaxies, considering only those galaxies which significantly affect the lens model in the vicinity of the \MgII absorber galaxy and bright arcs (Figure~\ref{fig:image}). 
We ignore the perturbers which are farther away since they have negligible effect on the results of this work, and since we lack suitable constraints on the lensing potential for regions beyond the bright arcs.

The lens model is constructed using the Glafic \citep{2010Glafic} package, with the following mass components (listed in Tables~\ref{tab:Ra-dec-information} and \ref{tab:lens model parameters}). 
The cluster mass distribution is modelled as a NFW profile \citep{Navarro1997} at $z=0.43$, with the central dominant (cD) galaxy modelled as a singular isothermal ellipsoid (SIE). 
Another SIE profile at $z=0.43$ is added to model the effect of the perturber galaxy seen to the north of arc 2 (Figure~\ref{fig:image}). Although we are unable to confirm spectroscopically, the perturber is photometrically consistent with the cluster redshift, having similar colors to the cluster red sequence galaxies (e.g. HST F606W-F814W~$\approx1.1$). We note that its precise redshift does not affect the results of this work.
The absorber galaxy at $z=0.77$ is modeled as a singular isothermal sphere (SIS). The SIS is scaled and treated as existing in the cluster $z=0.43$ lens plane for purposes of optimizing the Glafic lens model, while it is rescaled to $z=0.77$ for purposes of lens reconstruction and analysis of circumgalactic \MgII absorption. 

The lens model is constrained by multiple images of individual star-forming regions within arc 2. The arc has a clear fold-image symmetry in its visual morphology, in which several individual regions can be identified \citep[as also discussed by][]{Dai2020}. We use three prominent regions spanning the extent of the arc (denoted as A, B, and C; multiple images are labeled as A1, A2, A3, etc. in Figure~\ref{fig:critical_curve}). Further multiple images (C1, C2, C3) are seen around the perturber galaxy located to the north of arc 2 (Figure~\ref{fig:image}) along with a counter-image C4. This multiplicity of region C provides good lens model constraints on the perturber. Collectively these regions pinpoint the location of the critical curve through arc 2. 
\comment{ We also include the counter-image of arc 2 (images A3, B3, and C5 in Figure~\ref{fig:critical_curve}) as a constraint on the lens model. This counter image was initially identified based on consistent color and surface brightness, and we spectroscopically confirmed its nature as a multiple image of arc 2 with our 2020 data. 
}

We fit the lens model allowing all NFW parameters to vary, with a constraint on ellipticity to prevent overfitting of the model. Priors are placed on the SIE and SIS profiles to best fit the constraints.  Glafic determines the best-fit parameters using a downhill simplex method to find the region of minimal $\chi^2$. The values of the best-fit model are presented in Table \ref{tab:lens model parameters} along with the adopted priors. 
We note that the NFW profile mass is in good agreement with expectations based on the cluster velocity dispersion \citep{Bayliss2011}. 
Figure~\ref{fig:critical_curve} shows the location of the critical curve and predicted image positions for the best-fit model.

The southern giant arc (arc 1) was not used in modeling of this system and thus offers a key test of the lens model. Our spectroscopic data and the HST imaging indicate that the arc is a single highly magnified image, which our lens model accurately reproduces. We also note that our lens model produces the same general features as the model of \citet{Dai2020}, although there are some differences in the orientation of the critical curve beyond the region of the giant arcs and \MgII absorber galaxy, where we lack strong lensing constraints. 

From the lens model we determine the magnification factor of the $z=0.77$ galaxy to be $\mu=3.1$, calculated as the average areal magnification within a 2.5\arcsec~box centered on the galaxy. We estimate the uncertainty to be approximately 10\% or $\pm0.3$ in $\mu$, corresponding to a spatial offset of 1 arcsecond. 
The magnification is reasonably precise since the lensing potential is well constrained from the two $z\simeq3$ arcs in this vicinity. 

To calculate impact parameters relevant for analysis of spatial CGM structure in our data, we use the lens model to ray-trace the position of each KCWI spaxel to the $z=0.77$ absorber plane. The locations of the absorber galaxy and two arcs in this plane are shown in Figure~\ref{fig:reconstruction}. The center of the absorber galaxy is defined as the point of maximum continuum flux (marked in Figure~\ref{fig:reconstruction} for both image and source plane) and the impact parameter is calculated for all spaxels as the radial distance from the galaxy center, in the $z=0.77$ plane. 
\comment{We estimate uncertainty by varying the location of the lensing critical curve and calculating the change in impact parameter in the $z=0.77$ plane. An offset of 1 arcsecond in the $z=2.92$ critical curve results in fractional changes of only $\lesssim3$\% (1 standard deviation) with $<1$\% change in the average values, for the absorber and arc regions shown in Figure~\ref{fig:reconstruction}. We view such an offset as rather conservative given the stringent constraints on the critical curve in this region, and thus estimate typical fractional uncertainties $\sigma \lesssim 3$\% in impact parameter due to the lens model. }

\comment{Figure~\ref{fig:spaxel_example} shows the orientation of the absorber galaxy major axis, as determined by isophote orientation in the reconstructed HST F160W image.} 
%\TJ{(If splitting Fig 3 into two separate figures, the reference for major axis figure may change.)}
The lens model indicates that both arcs subtend azimuthal angles in the $z=0.77$ plane near the minor axes of the absorber galaxy, where we may expect outflow signatures (if present) to be prominent \citep[e.g.,][]{Martin2019,LanMo2018}.

\section{Physical Properties} \label{sec:analysis}

\subsection{Systemic Redshift}

\begin{figure*}[htb!]
\centerline{
\includegraphics[width=\textwidth]{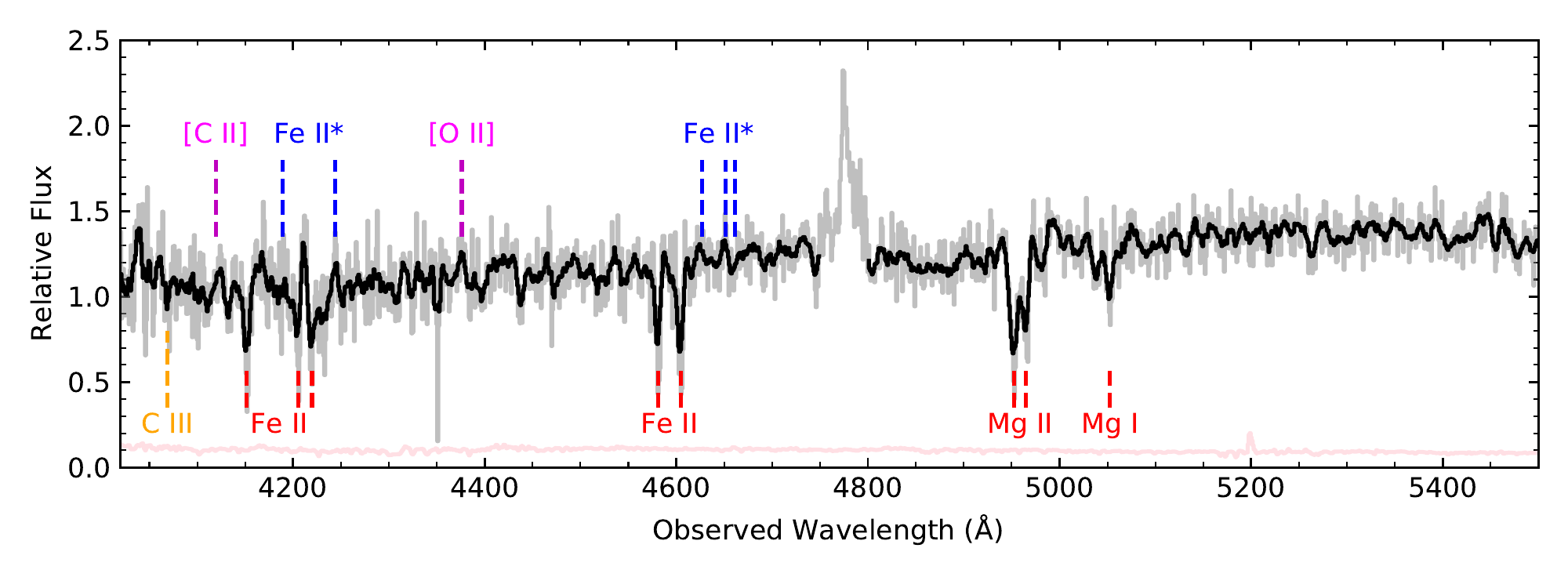}
}
\caption{
\textcolor{black}{KCWI spectrum of the absorber galaxy. The unsmoothed spectrum is shown in grey, while the black line is a running median over 7 pixels. The error spectrum is shown in pink. The prominent broad emission feature at $\sim$4770~\AA\ is scattered Ly$\alpha$ emission from the background $z\simeq2.9$ arcs. 
Notable spectral features of the absorber galaxy are labeled (for $z=0.771$). Color coding indicates the physical origin of each line (yellow: stellar photospheric absorption; red: interstellar absorption; blue: fluorescent fine structure emission; magenta: nebular emission). 
Interstellar absorption from \MgII, Mg~{\sc i}, and Fe~{\sc ii} are prominent in the spectrum. Fine structure Fe~{\sc ii}* and nebular [O~{\sc ii}] emission are weakly detected. }
}\label{fig:galaxy_spectrum}
\end{figure*}

\textcolor{black}{An accurate systemic redshift of the absorber galaxy is needed to assess the circumgalactic medium kinematics relative to the host galaxy. Figure~\ref{fig:galaxy_spectrum} shows the extracted KCWI spectrum of the absorber galaxy with prominent features labeled. The strongest lines are from interstellar absorption of \MgII and \FeII which are typically blueshifted relative to the stars. The most promising systemic features covered by the spectrum are weak nebular emission lines ([O\,II] and [C\,II]) and photospheric C\,III $\lambda$2997 absorption. Fine structure Fe\,II* fluorescent emission features can also be used to estimate the systemic redshift. }

\textcolor{black}{We fit Gaussian profiles to the emission features [O\,II]~$\lambda\lambda$2471, Fe\,II*~$\lambda$2396, and Fe\,II*~$\lambda$2626 in order to determine the systemic redshift. Each line is weakly detected at a S/N ratio of 4--5 (while the remaining nebular, fluorescent, and stellar features labeled in Figure~\ref{fig:galaxy_spectrum} are not significantly detected, with S/N~$<3$). We adopt the weighted mean best-fit redshift of these three lines: $z_{sys} = 0.77099 \pm 0.00011$. 
We note that the $\lambda$2396 and $\lambda$2626 lines are typically the strongest of the available Fe\,II* transitions, and are also those which most accurately trace the systemic velocity \citep[typically within 50~\kms; e.g.][]{Kornei2013}. Given that a systematic difference of $\sigma(z_{sys})\simeq0.0002$ is possible for the fine structure emission, and the [O\,II] fit has a redshift uncertainty $\sigma(z)=0.0004$, we caution that the adopted redshift uncertainty could be underestimated. 
Nonetheless all three lines are consistent with the mean $z_{sys}$ within their 1-$\sigma$ uncertainties, and within $\Delta z \leq 0.00013$ (or $<25$~\kms) in an absolute sense. }

\subsection{\MgII Absorption Line Kinematics}

Using the reduced KCWI data, we smoothed the flux measurements using a 2D gaussian filter of FWHM = 0.5\arcsec~to increase the signal-to-noise ratio (S/N) while preserving the spatial resolution (FWHM = 1.0\arcsec). From there we measured the absorption strength and kinematics of \MgII\,$\lambda\lambda2796, 2803$ at redshift $z=0.77$ in both arcs as well as the absorber galaxy. To quantify the \MgII absorption properties, we selected all spaxels corresponding to the absorber galaxy and the two background arcs with a minimum continuum \textcolor{black}{S/N~$>$~5$\sigma$} per spectral pixel, giving us a combined total of \textcolor{black}{280} selected spaxels. 

\begin{figure*}[htb!]

\centerline{
  \includegraphics[width=0.6\linewidth]{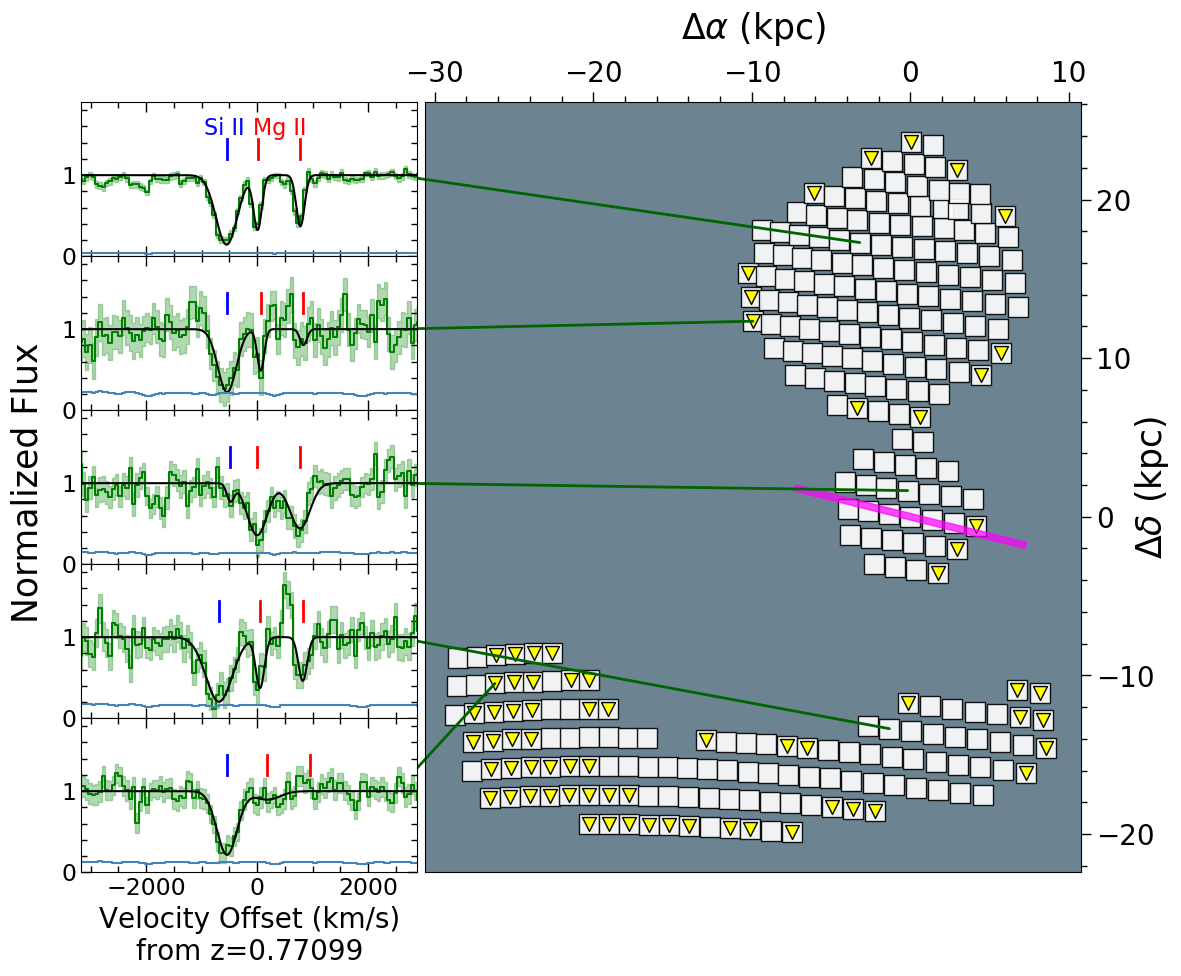}
}

\caption{
\comment{ 
 Spectra of individual spaxels showing absorption profiles from \MgII seen toward the absorber galaxy and background arcs (Section~\ref{sec:analysis}). The image on the right shows spaxel positions in the $z=0.77$ plane. Each spaxel shown has a continuum S/N $>5\sigma$ per spectral pixel near the \MgII absorption, while yellow triangles denote spaxels where \MgII absorption is not  detected ($<5\sigma$). The displayed spectra span a range of S/N in both the continuum and \MgII absorption, including \MgII detections and non-detections, illustrating the data quality. Raw spectra in each panel (green) are overlaid with a triple Gaussian fit (black) corresponding to the three most prominent absorption features: \SiII $\lambda$1260 in the background arcs at $z=2.92$, and the \MgII $\lambda\lambda$2796,2803 doublet at $z=0.77$ which is the main subject of this work. The error spectrum in each spaxel is shown in blue. The pink line denotes the morphological major axis of the absorber galaxy in the $z=0.77$ plane; this axis has an orientation angle of $\theta\approx76^\circ$.}
} \label{fig:spaxel_example}
\end{figure*}

\begin{figure*}[htb!]
\centerline{\hspace{-2em}
\includegraphics[width=1.05\linewidth]{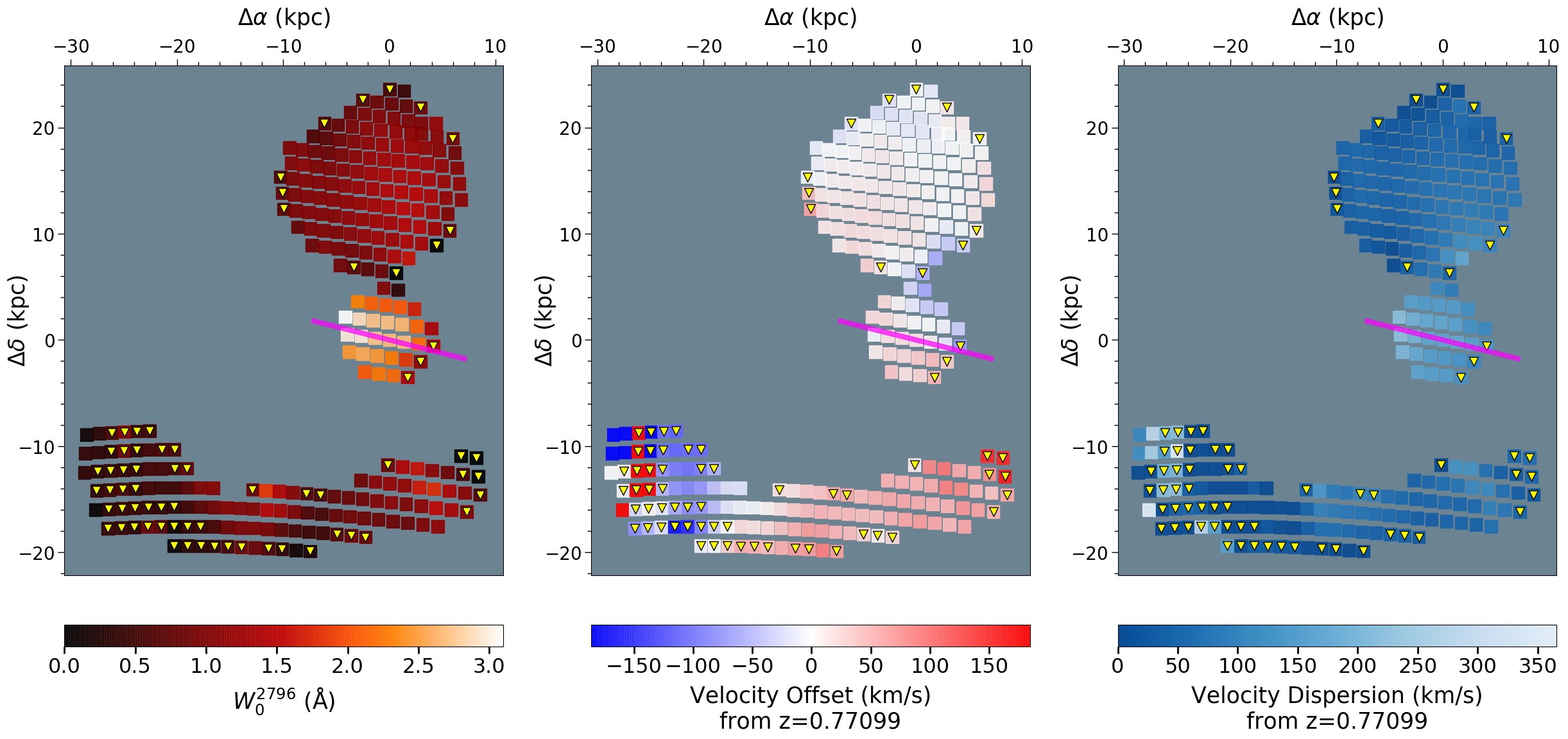}
}
\caption{
Colormaps of \MgII $\lambda2796$ rest-frame equivalent widths ({\em Left}), velocity offsets ({\em Middle}), and velocity dispersions ({\em Right}) of the \MgII gas in the z\,=\,0.77 absorbing plane of the galaxy and the background arcs. \textcolor{black}{Spaxels with undetected \MgII absorption ($\Sigma_{\text{Mg}}<5$; Table \ref{tab:full_properties}) are denoted by yellow downward arrows.} {\em Note:} The spaxels in the colormaps are distinguished by their offset separation $(\Delta\alpha, \Delta\delta)$ with respect to the center of the absorber galaxy (see Figure \ref{fig:reconstruction}).
} \label{fig:colormaps}
\end{figure*}

At each spaxel position, the spectrum near 4950\,\AA\ was fit using a sum of three gaussian profiles corresponding to the \MgII doublet at $z=0.77$ and the \SiII $\lambda$1260 absorption line at $z=2.92$ from the background arcs. The triple-gaussian fit is parameterized by the rest-frame equivalent width ($W_0$), velocity offset ($v$), and velocity dispersion ($\sigma$) for the three absorption lines\textcolor{black}{, allowing us to characterize the spatial trends of these parameters. Figure \ref{fig:spaxel_example} provides examples of the spectral fits for various spaxels within the absorber galaxy and both gravitational arcs. It is important to note that the models adopted a lower limit to the velocity dispersion given by the spectral resolution of KCWI (FWHM $= 2.4$\,\AA). The best-fit line widths were then corrected for this instrumental resolution to give us the intrinsic velocity dispersion measurements for each spaxel. Our fits used a common redshift and velocity dispersion for both \MgII lines in each spaxel, providing more robust fits and minimizing spurious fits to noise in low S/N regions.} Spatial maps of the best-fit equivalent widths, velocity offsets, and dispersions of \MgII absorption are shown in Figure~\ref{fig:colormaps}.

\textcolor{black}{Of the spaxels that have continuum S/N~$>$~5$\sigma$, we check the detection of \MgII absorption within the fit profiles using a chi-square test. We compare how the addition of \MgII components in our fit improves the $\chi^2$ value, compared to a fit with only the \SiII absorption, and thereby determine the detection significance $\Sigma_{\text{Mg}}$ of \MgII absorption in units of standard deviations. Spaxels whose \MgII absorption detection significance is marginal ($<5\sigma$) are considered non-detections and will be denoted in plots as either yellow triangles (e.g., Figures \ref{fig:colormaps} and \ref{fig:opacity_ratios}) or open-faced data points (e.g., Figure \ref{fig:impact_params}).} A full list of the fitted parameters, \MgII significance values ($\Sigma_{\text{Mg}}$), and continuum signal-to-noise ratios ($S/N$) for each spaxel is in Table~\ref{tab:full_properties}. The table also contains the offset separation ($\Delta\alpha$, $\Delta\delta$) of each spaxel with respect to the galaxy in the $z=0.77$ absorber plane (kpc) as well as the impact parameter of each spaxel in the absorber plane ($D$).

The colormaps show strong \MgII absorption with $W_0^{2796}>1$\,\AA~across a large area in both arcs, indicating the widespread presence of cool, metal-enriched circumgalactic gas. In addition, the absorption strength in the arcs is not uniform, indicating a clumpy medium similar to the findings of \citet{Lopez2018}, who also performed a similar tomographic observation a $z=0.98$ galaxy system (with somewhat lower inferred halo mass $M_{\text{halo}}\sim 10^{11}\,M_{\odot}$; cf. section \ref{sec:halo_mass} below) at impact parameters $\approx15$--90\,kpc. The \MgII absorption fits from larger impact parameters of arc 1 suggest inhomogeneities within the CGM. It is likely that the CGM can span the entirety of arc 1; however, the distribution of \MgII that is well-detected is relatively close (within $\sim$40\,kpc) to the absorber galaxy. While strong \MgII absorption is prominently seen from the CGM, the gas in the arcs shows little bulk motion relative to the galaxy: at first glance, the velocity offsets relative to the central galaxy are fairly small ($|v|\lesssim$ 80~\kms) and typical velocity dispersions seen in the background arcs are only $\simeq 50$~\kms. Since we do not detect any stellar or nebular features to determine the systematic redshift of the galaxy, we caution that velocity offsets are converted with respect to the average \MgII absorption redshift of the galaxy-arc system ($z\approx0.7711$). We anticipate that the true systemic redshift is likely underestimated and could differ by up to $\sim50\,\kms$, based on the differences in velocity offsets between the two arcs. Relative velocities within the CGM are nonetheless unaffected. These ``kinematically cold'' arcs contrast with the larger velocity dispersion ($\sigma\approx170$~\kms) and blueshift observed toward the \MgII absorber galaxy itself (``down-the-barrel'' kinematics; see Table~\ref{tab:arc_properties}), which is presumably due to outflowing gas driven by star formation in the galaxy.

\subsection{Stellar Mass}

We derive the stellar mass and other stellar population properties from the integrated spectral energy distribution. We measure broad-band fluxes in several filters from observations with the Hubble Space Telescope (programs GO-12368, GO-15378) and Pan-STARRS \citep{Chambers2016}, summarized in Table~\ref{tab:properties}. To calculate magnitudes, we sum the flux within a 2.25\arcsec$\times$3.25\arcsec~aperture which captures the full spatial extent detected in HST imaging. This aperture does not capture extended flux in Pan-STARRS images due to seeing, so we subtract 0.27 from the Pan-STARRS magnitudes to match the HST photometry. We adopt a minimum systematic uncertainty of 2\% in photometric fluxes, although the true uncertainty is likely even higher \citep{Ilbert06}. 
Photometric measurements are then fit with the stellar population synthesis code FAST \citep{Kriek09}. We adopt \citet{BruzualCharlot03} spectral templates with a Chabrier IMF, solar metallicity, \citet{Calzetti00} dust attenuation curve, and an exponentially declining star formation history. The resulting best-fit stellar mass is $\text{log}(M_*/M_\odot)=9.6^{+0.2}_{-0.1}$, corrected for a lensing magnification factor $\mu=3.1\pm0.3$. 
Adopting a constant star formation history, the best fit is $\text{log}(M_*/M_\odot)=9.8\pm0.1$ and $\text{SFR}=10\pm5~\sfrunit$. 
The specific star formation rate is consistent with the ``main sequence'' of star-forming galaxies at $z\approx0.8$ \citep[$\approx10^{-9.0}$ yr$^{-1}$; e.g.,][]{Whitaker2014}. 

\begin{deluxetable}{lC}[b!]
\tablecaption{Photometry of the $z=0.77$ \MgII Absorber Galaxy
\label{tab:properties}}
\tablecolumns{2}
% \tablenum{1}
\tablewidth{0pt}
\tablehead{
\colhead{Filter} &
\colhead{AB magnitude}
}
\startdata
HST/ACS F606W  &  22.47\pm0.02  \\
HST/ACS F814W  &  21.45\pm0.02  \\
HST/WFC3-IR F110W  &  20.91\pm0.02  \\
HST/WFC3-IR F160W  &  20.56\pm0.02  \\
Pan-STARRS g  &  23.07\pm0.24  \\
Pan-STARRS r  &  22.45\pm0.22  \\
Pan-STARRS i  &  21.47\pm0.09  \\
Pan-STARRS z  &  21.65\pm0.14  \\
\enddata
\tablecomments{The photometry is used to determine the stellar mass and SFR.}
\end{deluxetable}

\subsection{Halo Mass and Circular Velocity}\label{sec:halo_mass}

% For reference
% Values uncorrected for magnification: 
% $\text{log}(\mu M_*/M_\odot)=10.1^{+0.2}_{-0.1}$ for declining SFH, $\text{log}(\mu M_*/M_\odot)=10.3\pm0.1$ and $\mu\times\text{SFR}=30\pm15~\sfrunit$for constant SFH.]}
%

The dark matter halo properties, and in particular the expected rotation curve, are important for interpreting measurements of the circumgalactic gas kinematics. 
We estimate the dark matter halo mass using the stellar-to-halo mass relation of \cite{Behroozi2013}. 
For the source (absorber galaxy) redshift and stellar mass, the expected halo virial mass and radius are $\text{log}(M_{\text{halo}}/M_\odot)=11.6\pm0.2$ and $R_{
\rm vir} = 115\pm20$~kpc (defined as $R_{\rm vir} = (3 M_{\rm halo} / 4\pi \times 200 \rho_c)^{1/3}$ with $\rho_c$ being the critical density at z=0.77), accounting for both uncertainty in the stellar mass and $\sim$0.1 dex scatter in the stellar-to-halo mass relation. 
The expected circular rotation velocity $v_c$ is relatively insensitive to halo mass \citep[e.g.,][]{Bullock2017}. Over the range of radii of interest here -- from $\sim$10 kpc to of order half the virial radius -- a halo mass $\text{log}(M_{\text{halo}}/M_\odot)=11.6\pm0.2$ corresponds to $v_c \approx 100\pm20$ \kms. For a purely dispersion-supported halo, the expected velocity dispersion is $\sigma_m = \frac{v_c}{\sqrt{2}} = 70\pm10~\kms$ assuming an isothermal profile.

\section{Spatial and kinematic structure of the CGM} \label{sec:source_plane}

\subsection{Optical Depth and Covering Fraction of \MgII}

In this section, we examine whether variation in \MgII equivalent width is caused by differences in the gas covering fraction, column density, or a combination. A key diagnostic is the optical depth of \MgII absorption revealed by the doublet ratio. If lower equivalent width is due to low gas column density, then we expect to see a larger ratio of $W_0^{2796} / W_0^{2803} > 1$ \textcolor{black}{in optically thin regions of the arcs}, whereas if it is due to a lower covering fraction we expect a ratio closer to one \textcolor{black}{in the regions of low equivalent width}.

\begin{figure}[htb!]
    \centering
    \hspace{-2em}
    \includegraphics[width=\columnwidth]{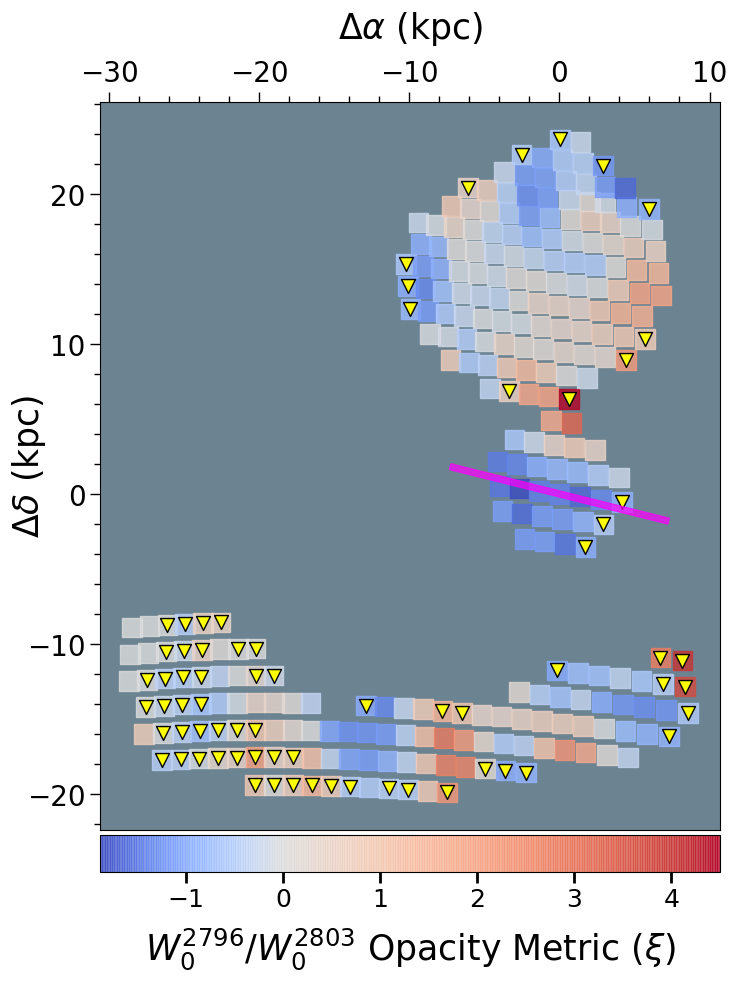}
    \caption{Colormap of the opacity metric $(\xi)$ of $W_0^{2796}/W_0^{2803}$ in the absorber plane in units of standard deviations. Values $\sim$\,0 correspond to optically thick gas whereas values $\lesssim \, -1$ and $\gtrsim \, 1$ may indicate optically thin gas and nonphysical absorption ratios, respectively.}
    \label{fig:opacity_ratios}
\end{figure}

Figure \ref{fig:opacity_ratios} illustrates the equivalent width line ratios of \MgII $\lambda2796$ $\left(W_0^{2796}\right)$ and \MgII $\lambda2803$ $\left(W_0^{2803}\right)$. The ratios are displayed in terms of an ``opacity metric'' $\xi$, where
\begin{equation}
    \xi = \frac{1-\frac{W_0^{2796}}{W_0^{2803}}}{\text{SD}\left(\frac{W_0^{2796}}{W_0^{2803}}\right)},
\end{equation}
where SD$(x)$ is the standard deviation of the line ratio $x$ obtained by propagating uncertainties. Optically thick gas is characterized by $W_0^{2796} \simeq W_0^{2803}$, so our metric classifies optically thick gas at $\xi=0$. Optically thin gas corresponds to values $\xi<0$; for example, a value of $\xi=-3$ would indicate that the \MgII absorption is not optically thick at 3$\sigma$ significance. Values $\xi>0$ are nonphysical for pure absorption, although in principle such values can arise from saturated absorption combined with emission line filling. 

The results show that the two arcs mostly vary within $\pm1\sigma$ of $\xi=0$, which is indicative of optically thick gas. We see minimal indication of optically thin gas, and nonphysical values  are located in spaxels near the edges of the arcs where the absorption signal is not as strong, possibly resulting from spurious fits. The low variation among the line ratios indicates that the low equivalent widths are driven primarily by kinematics and spatial covering fraction, rather than column density. Therefore, the variation in equivalent width appears to indicate a patchy spatial distribution of \MgII gas, similar to the results from \cite{Lopez2018,Lopez2020}.

\subsection{Equivalent Width vs. Impact Parameter}\label{sec:ew_b}

%\textcolor{blue}{**PLEASE REVISE AFTER SECTION 6**}

\begin{figure*}[htb!]
\centerline{
  \includegraphics[width=0.53\linewidth]{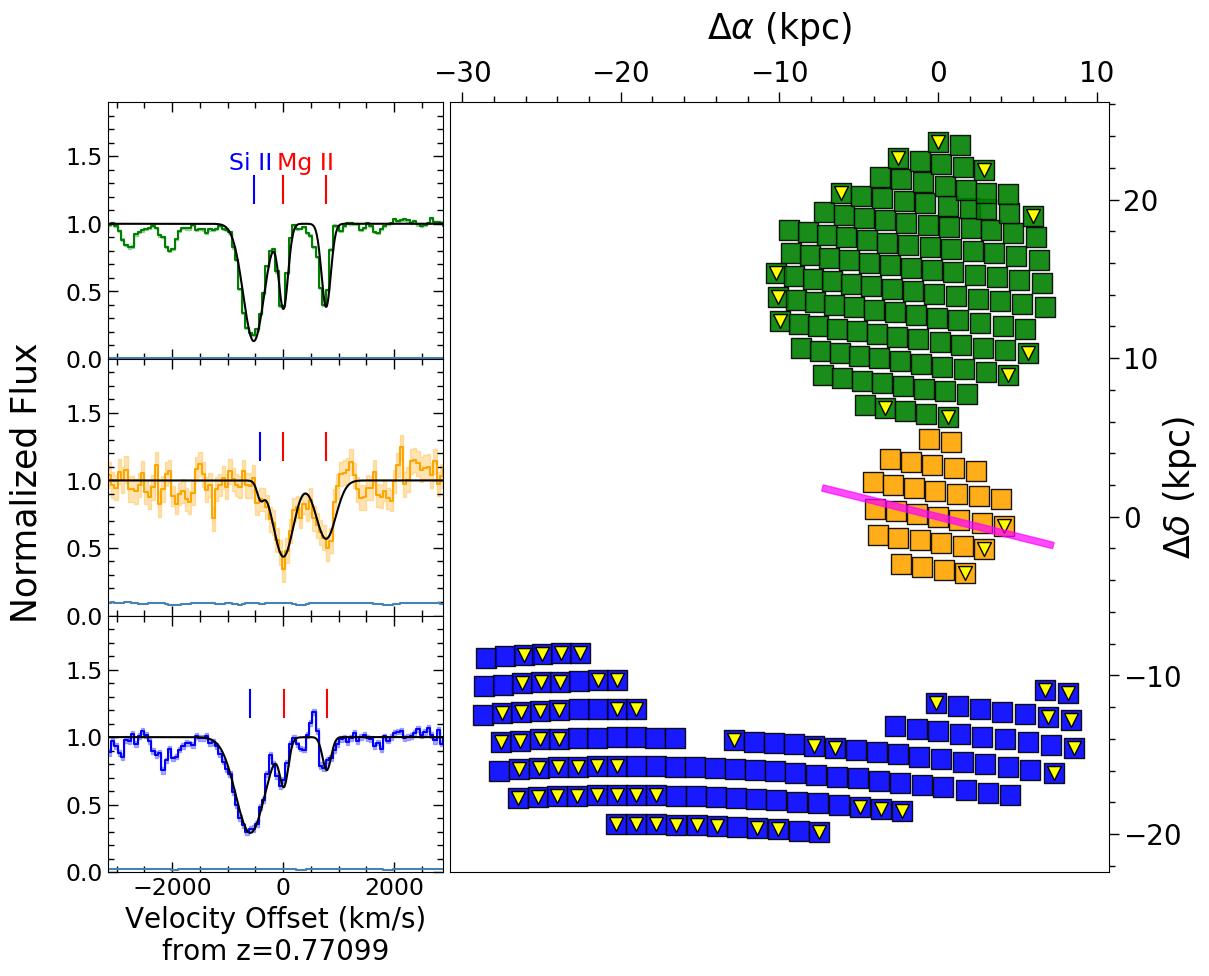}
  }
\vspace{0.25cm}
\centerline{
  \includegraphics[width=0.53\linewidth]{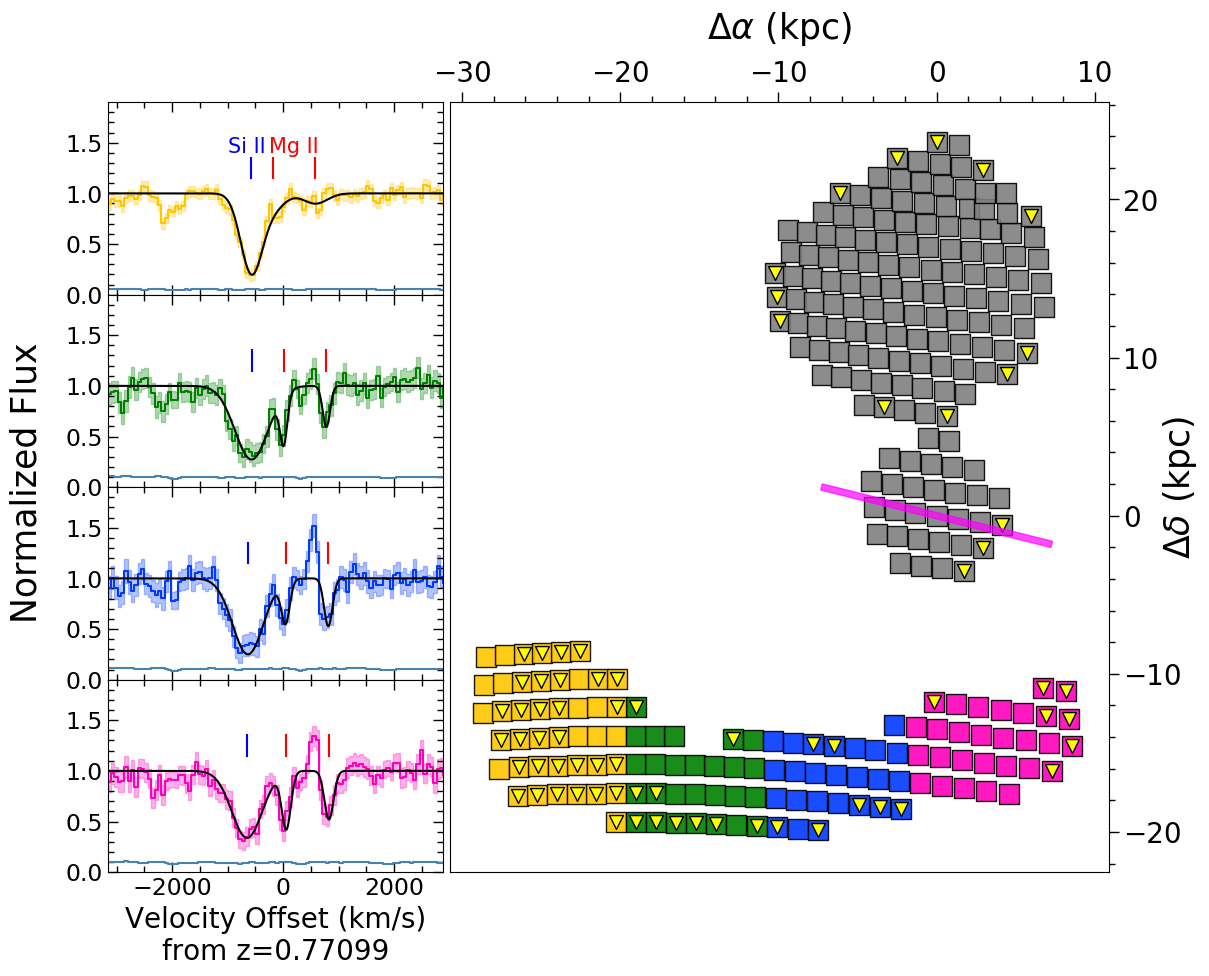}
  \hspace{0.25cm}
  \includegraphics[width=0.53\linewidth]{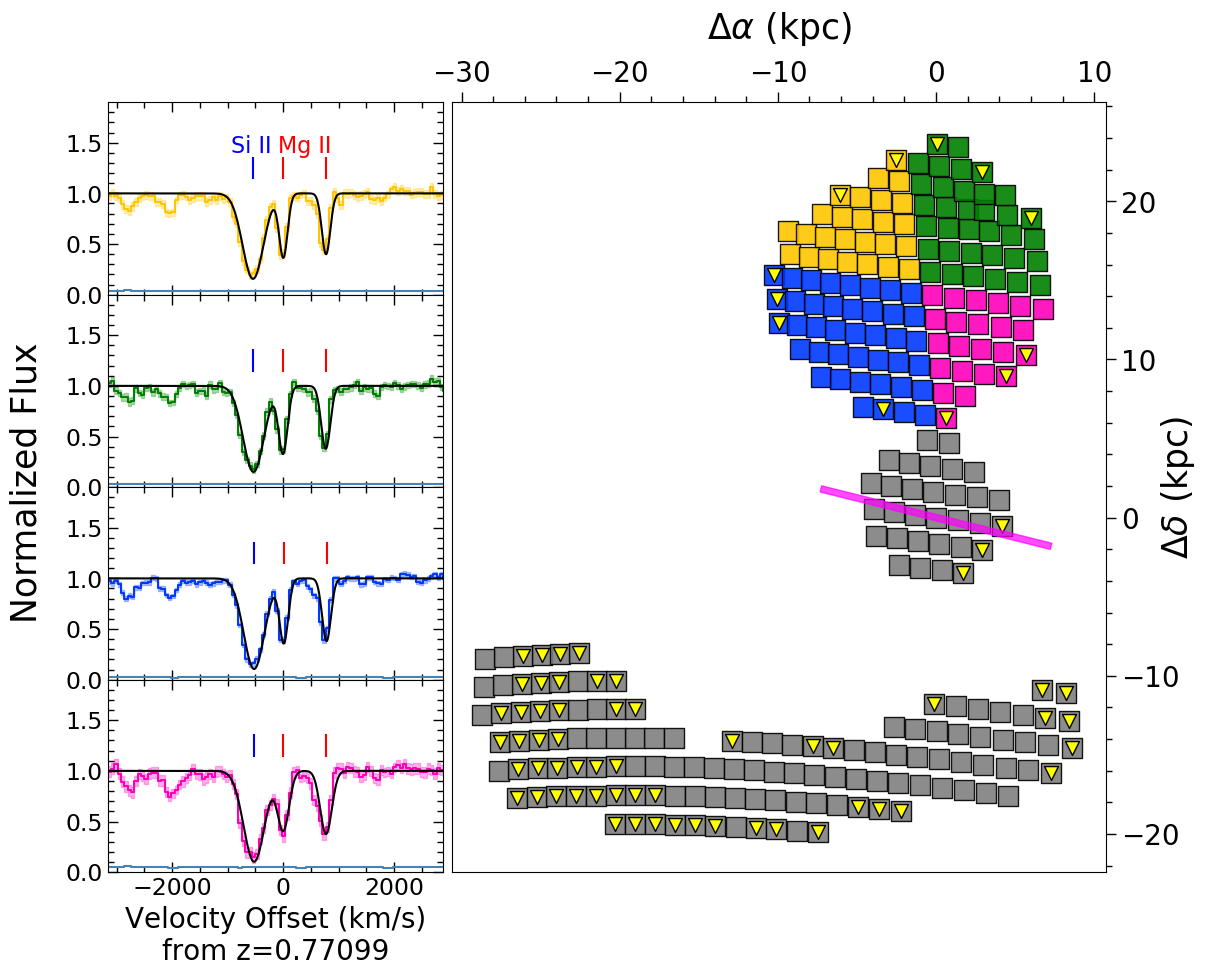}
}

\caption{
 \textcolor{black}{{\em Top:} Spatially integrated spectra of the absorber galaxy (orange) and background arcs (blue and green), showing the high S/N absorption signal obtained from summing all spaxels within each object. The background arcs show prominent \SiII $\lambda$1260 absorption at the arc redshifts ($z\simeq2.92$), unrelated to the \MgII absorption associated with the $z=0.77$ galaxy. The spatial regions corresponding to these integrated spectra are shown at right in the $z=0.77$ plane. 
 {\em Bottom:} Similar spectra from binned spaxels for different regions of arc 1 ({\em left}) and arc 2 ({\em right}), each separated into four sections of comparable area. The best-fit measurements for \MgII $\lambda2796$ rest-frame $W_0$, $v$, and $\sigma$ for these spatially binned regions, as well as the average impact parameter $D$, are reported in Table \ref{tab:arc_properties}.
 %\TJ{(Are the spaxels really for continuum S/N $> 3$? It looks like the same spaxels used in other figures which I thought were for S/N $> 5$.)}
 }
} \label{fig:spaxel_analysis}
\end{figure*}

To achieve a better understanding of the spatial distribution of \MgII gas in Figure \ref{fig:colormaps}, we compared the absorption profiles between the absorber galaxy and the two gravitational arcs through both a stacked spaxel analysis  (Figure~\ref{fig:spaxel_analysis} and Table \ref{tab:arc_properties}) and as a function of impact parameter $D$ (Figure~\ref{fig:impact_params}). \textcolor{black}{We caution that individual spaxels are not independent due to the seeing, causing the correlated patterns seen in arc 2 measurements. The true spatial resolution element corresponds to $\simeq10$ spaxels ($\simeq15$~kpc$^2$) and spans $\sim$2--8 kpc in Figure~\ref{fig:impact_params}. To account for the true resolution, we provided local regression curves to both arcs in Figure \ref{fig:impact_params} that were obtained through locally weighted scatterplot smoothing (LOWESS) methods. These LOWESS curves show the overall trends of the \MgII gas in the arc sightlines, while smoothing over the correlated spaxel patterns caused by the seeing.}

\textcolor{black}{The results show a moderate $W_0$--$D$ anti-correlation in both arcs, with absorption falling below the detection threshold for individual spaxels at impact parameters $D\gtrsim25\,\text{kpc}$ from the center of the galaxy (see Figure \ref{fig:spaxel_analysis}), limiting our analysis of the \MgII distribution to within $\sim$10--25\,kpc. The LOWESS curves indicate that the average equivalent width of \MgII absorption seen in arc 1 is similar to that in arc 2 at fixed impact parameter, and both arcs exhibit lower equivalent widths than the absorber galaxy. If there was uniform homogeneity in the gas, we would expect similar measurements of \MgII gas on both sides of the CGM (probed by the two arcs). However, there are variations of equivalent width within arc 1 (about 2--3\,$\sigma$ at some impact parameters) that are not apparent in arc 2 at similar impact parameters, resulting in a inconsistency with a purely symmetric distribution of circumgalactic gas at a more detailed level than what occurs on average. This anisotropy is broadly consistent with \MgII distributions seen around other galaxies (e.g., \citealt{Lopez2018,Lopez2020}), and is perhaps unsurprising given that \MgII is observed to vary strongly on kpc scales from lensed quasar sightlines \citep[e.g.,][]{Ellison2004}, and as a function of azimuthal angle in composite samples \citep[e.g.,][]{Bordoloi2011}.}

\begin{deluxetable*}{l|CCCC}[htb!]
\tablecaption{\MgII $\lambda2796$ Absorption Distribution \& Kinematics (Figure \ref{fig:spaxel_analysis})
\label{tab:arc_properties}}
\tablecolumns{5}
\tablewidth{0pt}
\tablehead{
\colhead{} & \colhead{\textcolor{black}{$D$}} & \colhead{$W_0$} & \colhead{$v$} & \colhead{$\sigma$}\\
\colhead{} & \colhead{\textcolor{black}{(kpc)}} & \colhead{($\mathrm{\AA}$)} & \colhead{(km s$^{-1}$)} & \colhead{(km s$^{-1}$)}}
\startdata
    \textbf{Galaxy (Orange)} & \textcolor{black}{1.79\pm0.97} &  \textcolor{black}{2.35 \pm 0.20} &   \textcolor{black}{4.08 \pm 16.44} &  \textcolor{black}{167.30 \pm 14.84} \\
\hline
  \textbf{Arc 1 (Blue)} & \textcolor{black}{22.49\pm6.23} &  \textcolor{black}{0.58 \pm 0.10} &  \textcolor{black}{16.76 \pm 11.60} &   \textcolor{black}{45.98 \pm 19.74} \\
 \hline
 Section 1 (Yellow) & \textcolor{black}{29.35\pm2.68} &  \textcolor{black}{0.69 \pm 0.30} &  \textcolor{black}{-155.28 \pm 86.51} &  \textcolor{black}{199.33 \pm 50.51} \\
 Section 2 (Green) & \textcolor{black}{23.61\pm2.34} &  \textcolor{black}{0.75 \pm 0.15} &   \textcolor{black}{9.54 \pm 10.04} &   \textcolor{black}{24.92 \pm 28.11} \\
 Section 3 (Blue) & \textcolor{black}{18.07\pm2.22} &  \textcolor{black}{0.73 \pm 0.14} &    \textcolor{black}{43.55 \pm 11.68} &   \textcolor{black}{43.41 \pm 20.44} \\
 Section 4 (Pink) & \textcolor{black}{15.10\pm1.89} &  \textcolor{black}{1.07 \pm 0.14} &     \textcolor{black}{56.64 \pm 9.70} &   \textcolor{black}{53.85 \pm 14.79} \\
\hline
 \textbf{Arc 2 (Green)} & \textcolor{black}{12.91\pm4.43} &  \textcolor{black}{1.21 \pm 0.09} &    \textcolor{black}{5.63 \pm 5.27} &    \textcolor{black}{56.73 \pm 7.83} \\
\hline
Section 1 (Yellow) & \textcolor{black}{19.13\pm1.76} &  \textcolor{black}{1.11 \pm 0.09} &  \textcolor{black}{3.81 \pm 5.44} &  \textcolor{black}{44.26 \pm 9.38} \\
 Section 2 (Green) & \textcolor{black}{19.01\pm2.36} &  \textcolor{black}{1.26 \pm 0.09} &   \textcolor{black}{-1.62 \pm 4.98} &   \textcolor{black}{53.73 \pm 7.63} \\
 Section 3 (Blue) & \textcolor{black}{12.70\pm3.06} &  \textcolor{black}{1.16 \pm 0.09} &    \textcolor{black}{12.57 \pm 5.31} &   \textcolor{black}{49.22 \pm 8.56} \\
 Section 4 (Pink) & \textcolor{black}{11.54\pm2.28} &  \textcolor{black}{1.40 \pm 0.12} &     \textcolor{black}{3.22 \pm 7.42} &   \textcolor{black}{83.70 \pm 9.23} \\
\enddata
\end{deluxetable*}

We compare our arc data in Figure~\ref{fig:impact_params} with quasar sightlines from 182 intermediate redshift galaxies ($0.072\leq z\leq 1.120$) in the \MgII Absorber-Galaxy Catalog \citep{Nielsen2013a}, as well as tomographic measurements of two intermediate redshift galaxy systems at $z=0.98$ and $z=0.73$ (with $\text{log}(M_{\text{halo}}/M_\odot)\approx11.0$ and 11.6, respectively) described in \cite{Lopez2018,Lopez2020}. In general, our data are in agreement with the trend of the quasar statistics ($\log W_{r}(2796) = \alpha_{1}D + \alpha_{2}$, where $\alpha_{1} = -\,0.015 \pm 0.002$ and $\alpha_{2} = 0.27 \pm 0.11$; \citealt{Nielsen2013a}) and fall well within the spread of the quasar sightlines (RMSE $\approx0.66$). This result is consistent with other individual galaxy measurements from lensed arc tomography \citep{Lopez2018,Lopez2020}. 

\textcolor{black}{A striking feature of Figure~\ref{fig:impact_params} is how closely the arc tomography data track the average of QSO sightlines. Our measurements from the CSWA 38 system, as well as the two systems studied by \cite{Lopez2018,Lopez2020}, show much smaller scatter than the quasar samples. We consider two possible explanations for the difference in scatter. On one hand, scatter may arise from halo-to-halo variations in the CGM of different galaxies as traced by \MgII absorption. The extent and equivalent width of \MgII is correlated with global galaxy properties such as stellar mass, SFR, environment, and redshift \citep[e.g.,][]{Bordoloi2011,Nielsen2013a}. The three galaxies studied with arc tomography are similar in terms of global properties which may explain their relative consistency in \MgII absorption equivalent width. The QSO comparison sample in Figure~\ref{fig:impact_params} includes a broader range of galaxy properties which can explain at least some of the larger scatter. 
Another effect is that small-scale fluctuations {\it within} the CGM around individual galaxies can give rise to larger scatter towards the QSO sightlines. Indeed, lensed QSO systems reveal considerable variation in \MgII absorption on $\sim$kpc scales \citep[e.g.,][]{Ellison2004,Chen2014,Rubin2018c}. The arc tomography data has spatial resolution of order $\sim$15~kpc${^2}$, orders of magnitude larger than the effective area of QSO sightlines. The larger cross-sectional area should average over such small-scale fluctuations and result in lower scatter for the arc tomography. We would expect similarly decreased scatter for unlensed background galaxy sightlines, which likewise subtend a much larger cross section than QSOs. 
}

\textcolor{black}{There is ample evidence to suggest that the two effects discussed above -- bulk halo-to-halo variations of the CGM, and intra-halo fluctuations on scales smaller than the effective resolution elements -- are both relevant. Given the small available lens tomography sample, we do not attempt to distinguish their relative contributions here. However we note that the intra-halo effect offers an avenue for probing the coherence length scale of absorption within the CGM. Scatter in $W_{r}(2796)$ versus $D$ should anti-correlate with the cross-sectional size of the background source, with a dependence on the physical size of individual CGM absorption clouds. 
Characterizing this scatter for sources of different size (e.g., QSOs versus galaxies in Figure~\ref{fig:impact_params}) can therefore provide new information on spatial structure of the CGM around distant galaxies. 
}

\subsection{Angular Momentum vs. \textcolor{black}{Velocity Dispersion} Support of CGM Gas}\label{sec:kinematics}

%\textcolor{blue}{**PLEASE REVISE AFTER SECTION 6**}

Comparison of the arcs with the absorber galaxy (Figure~\ref{fig:spaxel_analysis} and Table~\ref{tab:arc_properties}) reveals obvious differences in gas kinematics. \textcolor{black}{The absorber galaxy notably exhibits a best-fit \MgII absorption velocity dispersion of $\sigma \simeq 167\,\kms$, which is roughly $3.5\times$ larger than in both arcs 1 and 2 ($\sigma\simeq50\,\kms$). This velocity dispersion largely drives the higher absorption equivalent widths seen toward the absorber galaxy. Moreover, the \MgII profile in the absorber galaxy is clearly skewed toward negative velocities, reaching $\simeq -500\,\kms$ as we discuss further in Section~\ref{sec:model}.}
 %\TJ{I wouldn't say there is a ``distinct blueshift'' -- in fact the centroids are consistent in Table 4. How about this instead: \\ 
  %The absorber galaxy notably exhibits a best-fit \MgII absorption velocity dispersion of $\sigma \simeq 167\,\kms$, which is roughly $3.5\times$ larger than in both arcs 1 and 2 ($\sigma\simeq50\,\kms$). This velocity dispersion largely drives the higher absorption equivalent widths seen toward the absorber galaxy. Moreover, the \MgII profile in the absorber galaxy is clearly skewed toward negative velocities, reaching $\simeq -500\,\kms$ as we discuss further in Section~\ref{sec:model}. Such a broad and blueshifted velocity range suggests a significant outflow... [connecting with the text below]}
Such a broad velocity range suggests a significant outflow component seen ``down-the-barrel'' of the absorber galaxy, yet we see \textcolor{black}{only modest evidence} of such broad outflow velocities in the background arcs even at small impact parameters ($\sim10$\,kpc). We note that while down-the-barrel measurements probe line-of-sight kinematics, the absorption can be dominated by dense gas close to (or within) the galaxy and the velocity measurements do not tell us how far the outflowing gas extends from the galaxy. We now examine the extent to which velocity gradients in the background arcs can be attributed to bulk rotational motion of the CGM. 

\begin{figure*}[htb!]
    \centering
    \hbox{\hspace{7.5em} \includegraphics[width=0.63\linewidth]{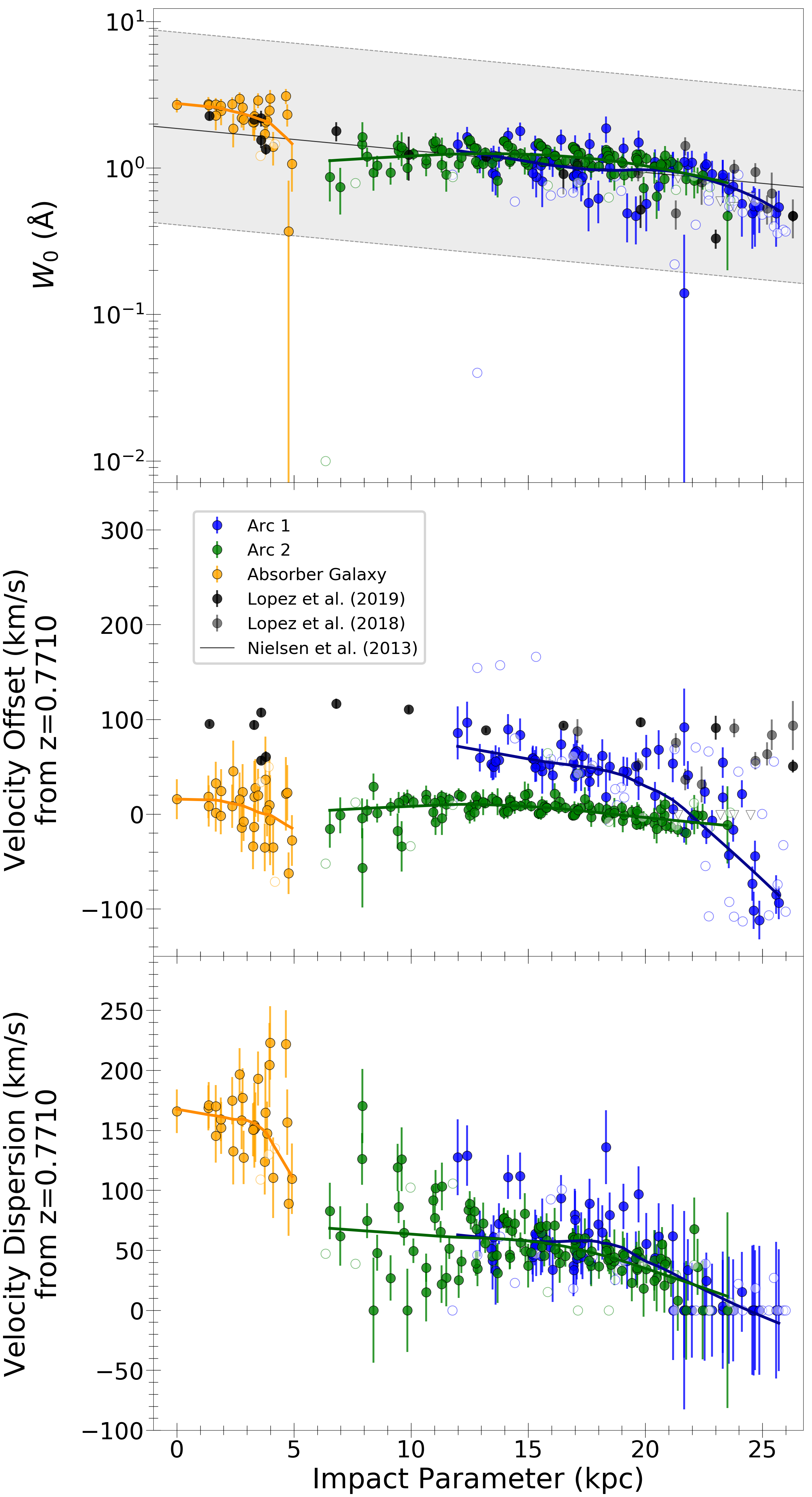}}
    \caption{\MgII $\lambda$2796 rest-frame equivalent width (\textit{Top}), velocity offset (\textit{Middle}), and velocity dispersion (\textit{Bottom}) as a function of impact parameter $D$ in the absorber plane \textcolor{black}{for arc 1 (blue), arc 2 (green), and the absorber galaxy (orange). Spaxels in each arc with undetected \MgII absorption ($\Sigma_{\text{Mg}}<5$) are denoted by open-faced data points of the equivalent colors in each plot. Error bars denote 1$\sigma$ uncertainty in the best-fit parameters of each spaxel.} The curves in each arc were obtained through locally weighted scatterplot smoothing (LOWESS) methods to understand the trends of the \MgII gas in the arcs while ignoring the correlated spaxel patterns affected by the seeing. For comparison, the black and grey data points are similar lensing tomographic measurements from \cite{Lopez2018,Lopez2020}. 
    %\textcolor{black}{and the red data points correspond to measurements from the MEGAFLOW survey of 22 quasar lines-of-sight \citep{Schroetter2019}}. 
    \textcolor{black}{In the top panel,} the black line and shaded region shows the maximum likelihood fit from \citet{Nielsen2013a} (for a sample of quasar-galaxy pairs) and the associated sample root-mean-square variation.}
    \label{fig:impact_params}
\end{figure*}

To further probe the velocity structure seen in the arcs, Figure~\ref{fig:impact_params} displays the velocity offsets and velocity dispersions as functions of impact parameter. \textcolor{black}{Our results indicate that the \MgII velocities show a modest variation within both arcs. The total range of velocity offsets spans $-110\lesssim v \lesssim$ 100\,\kms\ and averages $v \approx 15\pm15~\kms$. However, most spaxels at the low end of this velocity range are at impact parameters $\sim$25\,kpc, near the boundary of where we confidently detect \MgII absorption. The velocity gradient in arc 2 seen in Figure~\ref{fig:impact_params} is likely affected by spurious \MgII fits, as we do not see clear evidence of such a gradient in the spatially binned spectra (Figure~\ref{fig:spaxel_analysis}). Considering only robust detections at impact parameters $<20$~kpc, the total range of velocity offsets is only $\sim 0-100~\kms$. The range of best-fit velocity dispersions in the arcs is relatively large ($0\lesssim \sigma \lesssim$ 170\,\kms) although the majority of spaxels are near the average of $\sigma \approx 50\pm25~\kms$. Therefore the typical velocity FWHM ($\sim120~\kms$) in any given spaxel is comparable or larger than the variation in bulk motion seen across the entire system.} The mean $\sigma$ value is similar to the expected dark matter halo velocity dispersion of $\sigma_m = 70\pm10~\kms$ for a dispersion-supported system \citep[Section~\ref{sec:halo_mass};][]{Elahi2018}.

 %\TJ{I suggest revising the above sentence as: \\
  %The velocity gradient in arc 2 seen in Figure~\ref{fig:impact_params} is likely affected by spurious \MgII fits, as we do not see clear evidence of such a gradient in the spatially binned spectra (Figure~\ref{fig:spaxel_analysis}). Considering only robust detections at impact parameters $<20$~kpc, the total range of velocity offsets is only XXX [$\sim 0-100$ ish?]. } \\

Another important feature of the \MgII kinematics is that there is a noticeable difference in the velocity offsets of the arcs (Figure \ref{fig:impact_params}), which may provide evidence of angular momentum or a biconical outflow in the system. For an approximately isothermal density profile, we expect that the rotation $v_r$ and velocity dispersion $\sigma$ are related via $v_r^2 \approx v_c^2 - 2\sigma^2$ \citep[e.g.,][]{Burkert2010}. The bulk rotation velocity is expected to be lower than the circular velocity of the potential because the radial (turbulent) pressure gradient counteracts the centripetal acceleration \citep[see also][]{Wellons2019}. Here $v_c = 100\pm20~\kms$ is the circular velocity based on the estimated halo mass (Section~\ref{sec:halo_mass}), and we measure an average $\sigma=50~\kms$ from the background arc sightlines. We therefore expect the bulk rotation velocity to be $v_r=70\pm30~\kms$, with a ratio $v_r/\sigma = 1.4\pm0.6$. 
Our measured velocities imply smaller $v_r$ than this, but are compatible with this simple picture given the possible effects of inclination and orientation. \textcolor{black}{Since the lens model suggests that the arcs do not sample the kinematic major axis (Section~\ref{sec:lens_model}), the data do not robustly constrain the degree of rotational motion, nor the inclination or orientation of such possible rotation.} A rotation curve measurement for the absorber galaxy, which we do not have at present, would be valuable for further constraining the angular momentum.

The galaxies in \cite{Lopez2018,Lopez2020} exhibit higher velocity offsets compared to our data (Figure~\ref{fig:impact_params}), despite lower (but comparable) stellar masses of the host galaxies. 
This may further indicate that, in a relative sense, angular momentum is less important in the CGM of the CSWA 38 system studied here than in the systems studied by Lopez et al. In summary, our measurements of spatially resolved kinematics appear to be dominated by a relatively uniform velocity dispersion component, although we cannot rule out a substantial degree of rotation. 

\vspace{0.1cm}

\section{Physical Interpretation of the Circumgalactic Absorption} \label{sec:model}
%\textcolor{blue}{**PLEASE REVISE THIS SECTION FIRST**}

%Given the low velocity offsets and high velocity dispersions in both arcs (Section \ref{sec:kinematics}), we consider the case where CGM absorption arises from a \textcolor{black}{dispersion-supported} halo. 
%The halo mass of our absorber galaxy ($\sim4\times10^{11}~\Msun$) is roughly the mass scale above which the CGM is expected to become hot and pressure-supported \citep[e.g.,][]{BirnboimDekel2003, Keres2005, Faucher2011, Stern2019}, making this an interesting possibility.

\textcolor{black}{We now discuss the physical interpretation of velocity dispersions measured from single-component fits, which indicate typical $\sigma \approx 50$~\kms\ or FWHM~$\approx 120$~\kms\ throughout most regions of the background arcs. High-resolution spectroscopy of quasar sightlines reveals that low-ionization CGM absorption occurs in discrete clouds which individually have small Doppler $b$ parameters. Such clouds would be unresolved and blended at the spectral resolution of our KCWI data \citep[e.g.,][]{Zabl2020}. Moreover such clouds are likely far smaller than the effective $\sim$15~kpc$^2$ spatial resolution element of our data. Indeed, lensed quasar sightlines show that \MgII absorption profiles vary strongly on $\sim$kpc scales \citep[e.g.,][]{Ellison2004,Chen2014}. Therefore the absorption profiles measured for individual spaxels likely represent contributions from numerous discrete clouds, each of which is spectrally unresolved and has non-unity covering fraction even within a single resolution element. This explains the relatively smooth velocity and dispersion maps in Figure~\ref{fig:colormaps}, as we would expect more variation if only a small number of discrete absorbing clouds contributed to each spaxel (as discussed in Section~\ref{sec:ew_b}). 
Our kinematic measurements thus most likely represent the collective range of motions of individual clouds within the line-of-sight toward each spaxel (with minimal contribution from the clouds' intrinsic line widths). 
We therefore interpret these high velocity dispersions, as well as the low velocity offsets, as arising from dispersion support on the scales probed within the CGM.}

\textcolor{black}{Our argument for a mainly dispersion-supported CGM in our observed system differs from other tomographic surveys of gravitational arcs \citep{Lopez2018,Lopez2020}. The previous studies by Lopez et al. found evidence of a rotation signature in the CGM, suggesting that the observations probed accretion onto the absorber galaxy. However, as discussed in the previous section, our results favor a CGM supported largely by velocity dispersion. Since the galaxies studied by Lopez et al. are slightly less massive, and the inferred halo mass for our absorber galaxy near the value where simulations predict a CGM phase transition from cold to hot \citep[e.g.,][]{Keres2005,Stern2020}, the differences can plausibly arise in part from halo mass-dependent effects. 
Our results highlight that velocity dispersion, in addition to bulk velocity, is an important parameter for understanding CGM kinematics.}

\begin{figure*}[htb!]
\begin{tabular}{cc}
    \includegraphics[width=0.475\linewidth]{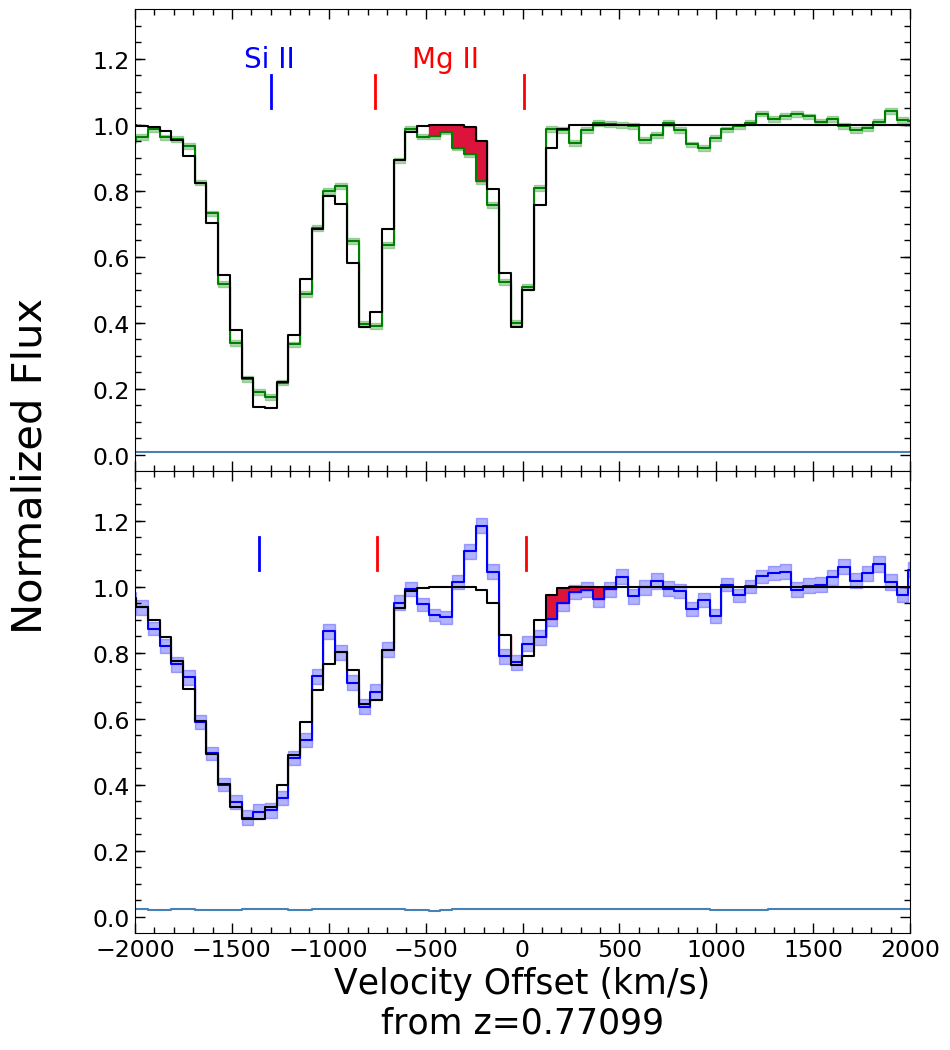} & \includegraphics[width=0.475\linewidth]{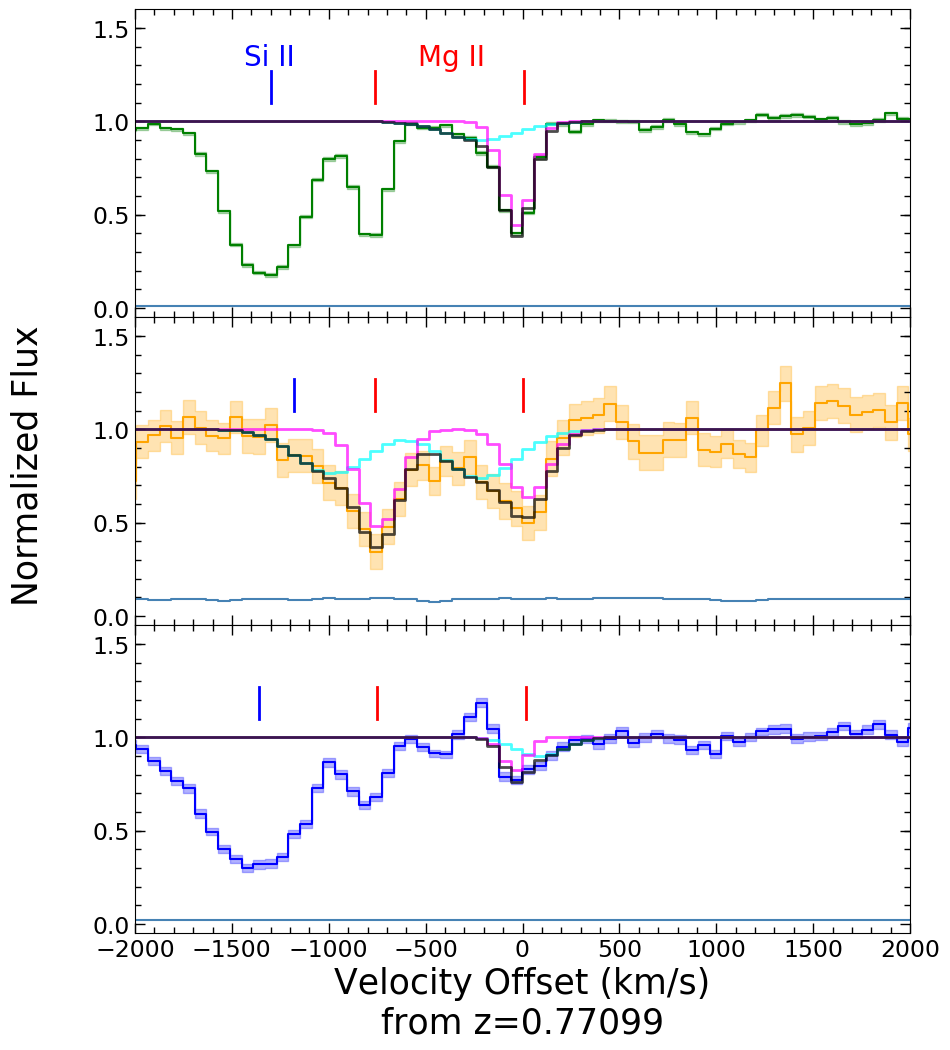}\\
    \Large{\qquad(a)} & \Large{\qquad(b)}
  \end{tabular}
\caption{
\textcolor{black}{(\textbf{a}) Triple-gaussian profiles fitted to \MgII absorption lines of arc 2 (\textit{top}) and arc 1 (\textit{bottom}). Red shaded areas reveal underestimations in the absorption strength fit profiles, indicating the possible presence of a second kinematic component. (\textbf{b}) Two-component gaussian fits (black) to \MgII absorption lines of arc 2 (\textit{top}), the absorber galaxy (\textit{middle}), and arc 1 (\textit{bottom}). Magenta profiles indicate narrow \textcolor{black}{dispersion-supported} gas components and cyan profiles indicate broad outflow components. In all cases the spectra are spatially integrated for the arcs and absorber galaxy. 
The velocity axes in this figure are centered on \MgII $\lambda$2803 and not \MgII $\lambda$2796; we do not fit the $\lambda$2796 line in the background arcs due to blending with the arcs' strong \SiII $\lambda$1260 feature.}
} \label{fig:two_comp}
\end{figure*}

We also consider the possibility of an underlying subdominant outflow component contributing to the \textcolor{black}{dispersion support} in the CGM. Our triple-gaussian fits for both arcs showed apparent underestimates in the cleaner \MgII $\lambda$2803 line (Figure~\ref{fig:two_comp}a). The residual absorption is at redshifted velocities in arc 1 and blueshifted in arc 2, consistent with the distinct velocity offset between the arcs (Figure \ref{fig:impact_params}). If this residual absorption were to originate from a bi-conical outflow component in the CGM, we would expect an improvement in our fits by searching for an underlying broad secondary component in addition to the dominant $\sigma\approx50~\kms$ CGM component. To test whether such an additional kinematic component is present, we fit the redder \MgII line with a double gaussian profile: a narrow component corresponding to \textcolor{black}{dispersion-dominated} gas, and a broader outflow component. We do not consider the $\lambda$2796 line due to blending with strong \SiII $\lambda$1260 in the background arcs. We applied the fit to both arcs and compared the results with the galaxy in Figure \ref{fig:two_comp}b.

\textcolor{black}{The absorber galaxy has a significant improvement with the addition of the outflow component in its down-the-barrel spectrum (corresponding to 5.5$\sigma$ detection of a second component), illustrating that the absorption profile is intrinsically non-gaussian. This is consistent with previous observations of outflows driven by stellar feedback from galactic disks, where down-the-barrel absorption profiles are typically asymmetric with a broad tail toward blueshifted velocities \citep[e.g.,][]{Bouche2012, Bordoloi2014, Rubin2014, Schroetter2016}. 
The narrow component of the fit is centered near the estimated systemic velocity and may represent the galaxy's interstellar medium, while the broader blueshifted component is clearly associated with outflowing gas detected out to roughly $-500$~\kms. 
The arcs, on the other hand, are characterized by a dominant narrow absorption profile with relatively little contribution from a broader outflow component. Arc 2 does experience a significant improvement when an outflow component is added (corresponding to 4.4$\sigma$ detection of a second component), whereas arc 1 shows no detectable improvement (0.6$\sigma$). 
The lack of outflow detection in arc 1 may be due in part to the lower S/N of its spectrum compared to arc 2, and the generally lower \MgII equivalent width corresponding to larger impact parameters. 
We conclude that there is statistically significant evidence of outflowing gas along the line-of-sight in at least some directions toward the background arcs, although the total absorption profile in the arcs is dominated by a $\sigma \approx 50$~\kms\ component centered near the systemic velocity. 
} 

\textcolor{black}{To determine the extent to which the outflow is detected in arc 2, we separately examined two halves of the arc corresponding to impact parameters of $\sim$12\,kpc and $\sim$19\,kpc. We found that a similar outflow signature was present at both impact parameters (with 5.0$\sigma$ and 3.0$\sigma$ significance, respectively). Therefore, it appears as though the outflow extends throughout the entirety of arc 2. 
However, the outflow is overpowered by the \textcolor{black}{dispersion-supported} CGM. %Interestingly, the subdued outflow in the CGM is broadly consistent with the FIRE galaxy formation simulations, which predict that as a hot, \textcolor{black}{dispersion-supported} CGM develops, outflows within the CGM are suppressed \citep{Muratov2015, Stern2020}. 
The presence of a prominent metal-enriched outflow in down-the-barrel absorption may trace ``recycling'' winds at relatively small impact parameters from the central galaxy \citep[e.g.,][]{Oppenheimer2010, Hafen2019b}.}

It is interesting to consider effects of azimuthal sampling. The gravitational lens model reconstruction (Section~\ref{sec:lens_model} and Figure~\ref{fig:reconstruction}) indicates that the two arcs are roughly aligned with the minor axes of the absorber galaxy, and do not sample along the major axis (Figure \ref{fig:spaxel_example}). Other studies have examined CGM absorption dependence on azimuthal angle \citep{Bordoloi2011, LanMo2018, Martin2019}, with typically large equivalent widths toward the minor axis where we expect gas outflows to be more prominent in star-forming disk galaxies \citep{Veilleux2005}. 
If the strong absorption we see along the minor axis is indicative of outflows, as other works suggest \citep[e.g.,][]{Bordoloi2011, Bouche2012, Kacprzak2012, Lan2014, LanMo2018}, one would expect the outflowing gas to have a velocity offset and a possibly large velocity dispersion. \textcolor{black}{While the arcs show a relatively small velocity offset with high dispersions, the velocity dispersions in the arcs overall are much smaller than the clear outflow component seen in the absorber galaxy spectrum (Figure \ref{fig:spaxel_analysis}; Table \ref{tab:arc_properties}), indicative of a subdominant outflowing gas component consistent with the two-component analysis discussed earlier. }

\textcolor{black}{To examine the extent to which this secondary outflow varies azimuthally, we performed the same two-component analysis on the sections of arc 2 previously configured in Figure \ref{fig:spaxel_analysis}. The analysis indicates that the outflow component appears fairly consistent throughout the four sections (2.6$\sigma$, 2.6$\sigma$, 5.2$\sigma$, 3.8$\sigma$ significance, respectively; see Table \ref{tab:arc_properties} for reference), with loss of significance likely contributed by the radial variation rather than azimuthal variation. 
Although we formally detect this outflow component in only one arc, we reiterate that arc 1 corresponds to larger impact parameters and has lower continuum S/N. Based on the regions of arc 2 with comparable $D$, we would expect $\lesssim 1.5\sigma$ significance of a comparable broad component for each binned region of arc 1 (as shown in Figure~\ref{fig:spaxel_analysis}), such that the non-detections are not constraining. We thus view the broad component in arc 2 as likely indicative of a (subdominant) biconical outflow of metal-enriched gas into the CGM, with detection across the azimuthal extent of arc 2 indicating at least a moderately wide opening angle.}

Some of the enriched CGM gas may be accreting onto the galaxy, rather than outflowing. The prevalence of cool, metal-enriched gas with modest velocity suggests that recycling gas replenishes the CGM, and may provide a reservoir to support ongoing star formation (possibly via spiraling inwards near the disk plane; e.g., \citealt{Ho2017, Martin2019}). 
However we cannot draw robust conclusions on inflows from the CGM based on our results. To confirm whether \MgII absorption may trace a bulk inflowing gas component, we would require measurements of the absorber galaxy rotation curve to model the expected CGM velocity field. Obtaining kinematic measurements of the absorber galaxy is therefore promising for further constraining inflows and outflows in this system.

\section{Discussion and Conclusions} \label{sec:discussion}

The CGM is important for understanding the gas flow processes that drive the evolution of galaxies. To better understand the CGM spatial structure and kinematics underlying these processes, we probed the cool metal-enriched CGM of a $z = 0.77$ star-forming galaxy (within the CSWA38 lens system) through a tomography technique using bright giant gravitational arcs as background sources. 
This adds to the currently small sample of galaxies whose CGM has been mapped spatially in absorption using integral field spectroscopy combined with gravitational lensing, a technique pioneered by \cite{Lopez2018}. 

Our study is based on observations obtained with the Keck Cosmic Web Imager. 
We have measured \MgII $\lambda\lambda$2796,2803 equivalent widths ($W_0$), velocity offsets ($v$), and velocity dispersions ($\sigma$) in a total of \textcolor{black}{280} spaxels, each corresponding to $\simeq$2 kpc$^2$ at $z=0.77$ for a typical magnification factor $\mu\simeq3$. The spatial resolution is 1.0\arcsec~FWHM in the image plane and $\simeq$15 kpc$^2$ at $z=0.77$. This configuration has allowed us to probe the CGM in an individual galactic environment at  %background 
impact parameters \textcolor{black}{$D \simeq 5$--30\,kpc}, in addition to the absorber galaxy using down-the-barrel spectroscopy ($D=0$). Our findings can be summarized as follows:

\begin{enumerate}
    \item CGM gas is well detected in \MgII absorption against both background arcs out to $D\approx25$\,kpc. Spatial variation in absorption equivalent widths combined with absorption line ratios indicate an optically thick medium with patchy distribution (i.e., varying covering fraction). These results are broadly consistent with the clumpy CGM inferred from previous tomographic CGM measurements using gravitational lensing \citep{Lopez2018,Lopez2020}.
    
    \item We observe a $W_0$--$D$ anti-correlation in the \MgII absorption. 
    \textcolor{black}{Both arcs in the CSWA~38 lens system lie near the mean of a fit to the \cite{Nielsen2013a} data,  
    %quasar log-linear maximum likelihood fit, 
    and the scatter of $W_0^{2796}$ in both arcs (Figure~\ref{fig:impact_params}) is far smaller than that measured from quasar sightlines through different halos. This result also holds for the two other systems with similar lens tomography measurements \citep{Lopez2018,Lopez2020}. 
    We attribute the relatively small scatter within lens tomography measurements to a combination of (1) ``halo-to-halo'' variation in the CGM around different galaxies, and (2) small-scale ``intra-halo'' variations within the CGM of individual galaxies. The latter point is especially interesting as it probes the size scale of absorbing clouds within the CGM. Since the arc tomography presented here effectively averages over areas of $\sim$15\,kpc$^2$ (i.e., much larger than for QSOs), the small scatter in the $W_0$--$D$ relation suggests that \MgII absorption in the CGM arises from individual components of $\lesssim$ kpc size, in qualitative agreement with intra-halo variations seen toward lensed QSO sightlines.}

    \item Absorption line kinematics in the arcs suggest that the \MgII bearing gas is largely \textcolor{black}{dispersion-supported in the regions probed}, in contrast to other systems which show a higher degree of rotational support \citep{Lopez2018,Lopez2020}. The velocity dispersion is at least half of the expected rotation velocity. \textcolor{black}{The absorber galaxy spectrum shows clear outflow kinematics in down-the-barrel \MgII absorption, while evidence of a subdued outflow component is prevalent in one of the background arc sightlines (arc 2) and is detected over $D \approx 10-20$\,kpc.}
\end{enumerate}

Future study of the CSWA38 lens system will benefit from an accurate measurement of the absorber galaxy rotation curve, for example from nebular emission lines such as \OII or H$\alpha$. This will either help confirm the \textcolor{black}{dispersion-dominated} interpretation or identify a more significant rotation component. 
Regardless, our study demonstrates that tomographic mapping continues to provide new, more detailed insights into the structure and kinematics of the CGM. 
This technique thus adds considerably to our toolkit for understanding the gas inflow/outflow processes that regulate star formation and quenching in evolving galaxies.

\section*{ACKNOWLEDGEMENTS}

This work is based on data obtained at the W. M. Keck Observatory, which is operated as a scientific partnership among the California Institute of Technology, the University of California, and the National Aeronautics and Space Administration. The Observatory was made possible by the generous financial support of the W. M. Keck Foundation. 
We wish to acknowledge the very significant cultural role and reverence that the summit of Maunakea has within the indigenous Hawaiian community. We are most fortunate to have the opportunity to conduct observations from this sacred mountain, and we respectfully say mahalo. 
\textcolor{black}{KM acknowledges support for this work from the National Science Foundation (NSF) under grant PHY-1852581 and the Dean's Distinguished Graduate Fellowship, funded by the Office of Graduate Studies and the Dean of the College of Letters and Science at the University of California, Davis.} TJ and KVGC acknowledge support from the Gordon and Betty Moore Foundation through grant GBMF8549. CAFG was supported by NSF through grants AST-1715216 and CAREER award AST-1652522, by NASA through grant 17-ATP17-0067, by STScI through grants HST-GO-14681.011, HST-GO-14268.022-A, HST-AR-14293.001-A, and HST-AR-16124.001-A, and by a Cottrell Scholar Award from the Research Corporation for Science Advancement. 
We thank the anonymous referee for a careful and constructive report which has improved this manuscript. 

%\TJ{(added a note acknowledging the referee)}


\vspace{5mm}
\facilities{Keck (KCWI)}

\bibliography{cswa38_mgii}
% \begin{thebibliography}{}
% \end{thebibliography}

\startlongtable
\begin{deluxetable*}{CCCCCCCCCC}
\tablecolumns{10}
\tablewidth{0pt}
\tablecaption{Extended Data Table of \MgII Absorption Distribution \& Kinematics \label{tab:full_properties}}
\tablehead{
\colhead{$\Delta\alpha$\tablenotemark{$a$}} & \colhead{$\Delta\delta$\tablenotemark{$a$}} & \colhead{$D$\tablenotemark{$b$}} & \colhead{$W_0$\tablenotemark{$c$}} & \colhead{$v$\tablenotemark{$d$}} & \colhead{$\sigma$\tablenotemark{$e$}} & \colhead{$\Sigma_{\text{Mg}}$\tablenotemark{$f$}} & \colhead{$S/N$\tablenotemark{$g$}}\\
\colhead{(kpc)} & 
\colhead{(kpc)} & 
\colhead{(kpc)} & 
\colhead{($\mathrm{\AA}$)} & 
\colhead{(km s$^{-1}$)} & 
\colhead{(km s$^{-1}$)}& 
\colhead{} & 
\colhead{}}
\startdata
 -20.27 &             -19.39 &              28.05 &    0.09\pm0.66 &     -8.62\pm174.97 &   157.38\pm173.04 &   1.22 &   5.11 \\
            -19.02 &             -19.39 &              27.16 &    0.35\pm0.27 &      -2.45\pm32.07 &       0.0\pm72.61 &   1.61 &   5.47 \\
            -17.75 &             -19.40 &              26.30 &    0.34\pm0.28 &      30.09\pm27.36 &      26.07\pm46.9 &   2.80 &   5.88 \\
            -16.49 &             -19.42 &              25.48 &     0.4\pm0.27 &      55.59\pm20.11 &     27.01\pm35.89 &   4.20 &   5.68 \\
            -15.22 &             -19.45 &              24.70 &    0.59\pm0.28 &      53.43\pm17.47 &     18.33\pm36.57 &   4.52 &   5.59 \\
            -13.95 &             -19.50 &              23.97 &      0.9\pm0.3 &      44.81\pm17.69 &     21.99\pm34.96 &   4.98 &   5.38 \\
            -12.67 &             -19.56 &              23.30 &    0.86\pm0.28 &      54.35\pm16.26 &       0.0\pm48.87 &   5.33 &   5.78 \\
            -11.38 &             -19.62 &              22.69 &    0.66\pm0.27 &      66.16\pm19.56 &       0.0\pm54.52 &   4.21 &   6.14 \\
            -10.09 &             -19.70 &              22.14 &    0.41\pm0.26 &      70.42\pm33.05 &        0.0\pm74.6 &   1.62 &   6.41 \\
             -8.79 &             -19.79 &              21.66 &    0.14\pm0.21 &       91.64\pm40.7 &       0.0\pm82.61 &    -- &   6.42 \\
             -7.47 &             -19.89 &              21.25 &    0.22\pm0.19 &      68.61\pm20.65 &       0.0\pm54.54 &   3.05 &   5.74 \\
            -26.49 &             -17.75 &              31.89 &    0.44\pm0.25 &     -80.05\pm36.43 &       0.0\pm77.75 &   0.93 &   5.05 \\
            -25.24 &             -17.69 &              30.82 &    0.46\pm0.23 &     -49.56\pm28.02 &       0.0\pm67.31 &   2.31 &   5.93 \\
            -24.00 &             -17.63 &              29.78 &    0.34\pm0.23 &     -20.05\pm30.82 &       0.0\pm70.84 &   1.97 &   6.58 \\
            -22.75 &             -17.59 &              28.76 &    0.85\pm1.01 &   -170.38\pm160.91 &    279.6\pm125.73 &   4.09 &   6.96 \\
            -21.51 &             -17.57 &              27.77 &    0.76\pm0.63 &    -156.59\pm75.75 &    160.88\pm48.77 &   4.50 &   6.84 \\
            -20.26 &             -17.55 &              26.81 &    0.11\pm0.22 &     -91.94\pm35.21 &       0.0\pm80.91 &   2.96 &   6.77 \\
            -19.02 &             -17.55 &              25.88 &    0.38\pm0.23 &     -32.61\pm21.39 &       0.0\pm57.99 &   3.43 &   6.97 \\
            -17.77 &             -17.56 &              24.98 &    0.48\pm0.23 &       0.15\pm17.59 &        0.0\pm50.9 &   4.75 &   7.11 \\
            -16.52 &             -17.58 &              24.13 &    0.57\pm0.23 &      21.37\pm14.45 &     15.41\pm33.31 &   5.99 &   6.72 \\
            -15.26 &             -17.62 &              23.31 &     0.9\pm0.23 &      17.74\pm12.37 &      3.19\pm38.77 &   6.53 &   6.47 \\
            -14.00 &             -17.67 &              22.54 &    1.03\pm0.23 &      23.71\pm12.66 &       0.0\pm42.12 &   6.52 &   6.49 \\
            -12.73 &             -17.72 &              21.82 &    1.05\pm0.24 &      41.18\pm14.75 &      33.6\pm25.54 &   6.65 &   6.91 \\
            -11.46 &             -17.79 &              21.17 &     1.0\pm0.28 &      53.33\pm19.69 &      61.7\pm26.76 &   6.29 &   7.41 \\
            -10.18 &             -17.88 &              20.57 &    0.75\pm0.24 &      68.04\pm20.59 &     61.42\pm27.81 &   5.25 &   7.84 \\
             -8.88 &             -17.97 &              20.04 &     0.57\pm0.2 &      65.01\pm18.63 &     55.34\pm25.61 &   5.24 &   7.93 \\
             -7.58 &             -18.07 &              19.60 &    0.47\pm0.17 &      50.55\pm14.63 &     43.37\pm22.29 &   5.97 &   7.38 \\
             -6.26 &             -18.18 &              19.23 &    0.49\pm0.18 &      44.67\pm15.05 &     41.16\pm23.35 &   5.68 &   6.63 \\
             -4.94 &             -18.30 &              18.96 &     0.7\pm0.21 &       28.3\pm17.24 &       36.1\pm28.0 &   4.68 &   5.99 \\
             -3.60 &             -18.43 &              18.78 &     0.9\pm0.26 &       9.52\pm19.77 &      34.1\pm32.55 &   4.17 &   5.53 \\
             -2.24 &             -18.57 &              18.71 &    0.88\pm0.26 &       26.44\pm20.4 &     25.49\pm36.31 &   3.73 &   5.26 \\
            -27.66 &             -16.00 &              31.95 &    0.06\pm0.86 &    184.63\pm562.58 &   365.39\pm521.57 &    -- &   5.74 \\
            -26.42 &             -15.92 &              30.85 &    0.39\pm0.19 &      -26.11\pm32.1 &       0.0\pm71.24 &   1.79 &   7.78 \\
            -25.18 &             -15.86 &              29.76 &    0.46\pm0.18 &       2.88\pm30.34 &      37.13\pm45.1 &   2.83 &   9.11 \\
            -23.95 &             -15.80 &              28.70 &    0.47\pm0.19 &     -13.71\pm31.77 &      48.6\pm43.97 &   3.25 &   9.54 \\
            -22.72 &             -15.76 &              27.65 &    0.44\pm0.22 &     -57.37\pm28.95 &      38.65\pm45.5 &   3.62 &   9.39 \\
            -21.49 &             -15.74 &              26.63 &    0.41\pm0.21 &     -80.62\pm22.64 &        0.0\pm61.5 &   4.23 &   8.81 \\
            -20.26 &             -15.72 &              25.64 &    0.36\pm0.21 &     -74.23\pm20.09 &       0.0\pm57.93 &   4.44 &   8.34 \\
            -19.02 &             -15.72 &              24.68 &    0.51\pm0.22 &     -44.47\pm16.27 &       0.0\pm49.91 &   5.29 &   7.76 \\
            -17.79 &             -15.73 &              23.75 &    0.75\pm0.22 &     -16.47\pm12.74 &       0.0\pm42.53 &   6.82 &   7.21 \\
            -16.55 &             -15.76 &              22.85 &    0.91\pm0.21 &      -6.76\pm10.92 &       0.0\pm38.84 &   7.33 &   6.51 \\
            -15.31 &             -15.79 &              21.99 &    1.09\pm0.22 &      -5.21\pm11.37 &        0.0\pm39.5 &   6.81 &   6.19 \\
            -14.06 &             -15.84 &              21.18 &    1.07\pm0.22 &        5.2\pm12.37 &       0.0\pm41.43 &   6.17 &   6.13 \\
            -12.80 &             -15.90 &              20.42 &     1.1\pm0.24 &      19.34\pm14.78 &     43.64\pm23.15 &   6.34 &   6.66 \\
            -11.54 &             -15.98 &              19.71 &      1.5\pm0.3 &        34.7\pm19.7 &     96.62\pm23.38 &   8.05 &   7.16 \\
            -10.26 &             -16.06 &              19.06 &    1.36\pm0.25 &      45.26\pm15.39 &     86.51\pm18.91 &   8.99 &   7.75 \\
             -8.98 &             -16.16 &              18.48 &    1.01\pm0.21 &      50.29\pm14.91 &     79.48\pm18.57 &   8.43 &   7.81 \\
             -7.69 &             -16.26 &              17.99 &     0.62\pm0.2 &      45.86\pm17.27 &      71.63\pm22.0 &   7.08 &   7.54 \\
             -6.38 &             -16.38 &              17.58 &    0.58\pm0.21 &      44.14\pm19.72 &     64.23\pm25.66 &   6.21 &   7.11 \\
             -5.06 &             -16.50 &              17.26 &    0.82\pm0.23 &      43.45\pm19.05 &     55.66\pm25.93 &   5.97 &   6.65 \\
             -3.73 &             -16.63 &              17.05 &    0.91\pm0.23 &      42.75\pm18.66 &     47.84\pm26.81 &   5.55 &   6.40 \\
             -2.39 &             -16.77 &              16.94 &    0.88\pm0.21 &      54.21\pm15.88 &     37.12\pm25.46 &   5.63 &   6.50 \\
             -1.03 &             -16.92 &              16.95 &    0.88\pm0.18 &       62.28\pm11.6 &      33.6\pm20.18 &   7.36 &   6.40 \\
              0.35 &             -17.07 &              17.08 &      0.9\pm0.2 &      66.59\pm11.12 &     43.07\pm17.39 &   8.87 &   6.60 \\
              1.74 &             -17.24 &              17.32 &    0.87\pm0.22 &      61.14\pm11.99 &     44.56\pm18.53 &   8.51 &   6.87 \\
              3.15 &             -17.40 &              17.68 &    1.04\pm0.21 &      47.72\pm11.58 &     40.26\pm18.85 &   7.73 &   6.72 \\
              4.57 &             -17.57 &              18.16 &    1.17\pm0.26 &      61.59\pm17.49 &     64.79\pm22.68 &   6.47 &   5.85 \\
            -27.57 &             -14.19 &              31.01 &     0.4\pm0.17 &       -7.74\pm30.8 &       0.0\pm68.87 &   1.64 &   7.47 \\
            -26.35 &             -14.11 &              29.89 &     0.5\pm0.31 &    184.63\pm151.11 &   221.04\pm149.03 &   2.96 &   9.91 \\
            -25.12 &             -14.04 &              28.78 &     0.6\pm0.33 &    184.63\pm141.18 &   226.64\pm138.12 &   3.88 &  11.52 \\
            -23.90 &             -13.99 &              27.70 &     0.4\pm0.16 &      -51.08\pm27.5 &     33.88\pm43.81 &   3.89 &  11.67 \\
            -22.69 &             -13.95 &              26.63 &    0.51\pm0.18 &     -75.89\pm19.85 &     20.27\pm39.87 &   5.27 &  10.88 \\
            -21.47 &             -13.92 &              25.59 &     0.49\pm0.2 &     -84.95\pm20.31 &       0.0\pm56.99 &   5.01 &   9.67 \\
            -20.25 &             -13.91 &              24.56 &    0.49\pm0.21 &      -73.39\pm18.1 &       0.0\pm53.93 &   5.01 &   8.43 \\
            -19.03 &             -13.91 &              23.57 &    0.71\pm0.22 &     -43.33\pm13.57 &       0.0\pm44.71 &   6.45 &   7.16 \\
            -17.81 &             -13.92 &              22.60 &    1.06\pm0.23 &     -20.19\pm11.38 &      24.47\pm23.5 &   7.96 &   6.28 \\
            -16.58 &             -13.94 &              21.66 &     1.1\pm0.23 &     -18.86\pm11.41 &      0.01\pm39.57 &   6.88 &   5.58 \\
            -12.87 &             -14.10 &              19.09 &    0.96\pm0.29 &      17.59\pm19.36 &      41.51\pm30.8 &   4.20 &   5.27 \\
            -11.62 &             -14.17 &              18.32 &    1.87\pm0.38 &      18.05\pm29.37 &    135.97\pm30.81 &   6.72 &   5.73 \\
            -10.35 &             -14.26 &              17.62 &    1.46\pm0.28 &      34.47\pm16.91 &      88.92\pm20.7 &   7.83 &   6.16 \\
             -9.08 &             -14.35 &              16.99 &     1.1\pm0.25 &      39.49\pm17.42 &     79.55\pm21.71 &   7.01 &   6.29 \\
             -7.80 &             -14.46 &              16.43 &    0.68\pm0.28 &      55.96\pm31.33 &    100.53\pm36.75 &   4.72 &   6.29 \\
             -6.50 &             -14.58 &              15.96 &     0.65\pm0.3 &      58.39\pm35.76 &     92.56\pm41.87 &   4.25 &   6.16 \\
             -5.19 &             -14.71 &              15.60 &    0.81\pm0.27 &      45.17\pm23.75 &     62.33\pm30.72 &   5.32 &   6.10 \\
             -3.87 &             -14.84 &              15.34 &    0.87\pm0.24 &      52.89\pm18.63 &     53.93\pm25.56 &   5.93 &   6.21 \\
             -2.53 &             -14.98 &              15.20 &     0.92\pm0.2 &      58.54\pm13.78 &     44.53\pm20.82 &   6.92 &   6.59 \\
             -1.18 &             -15.13 &              15.18 &    1.08\pm0.19 &      59.13\pm11.23 &     46.71\pm16.88 &   9.09 &   6.96 \\
              0.19 &             -15.29 &              15.29 &    1.16\pm0.19 &      57.09\pm10.65 &     52.69\pm15.29 &  10.50 &   7.47 \\
              1.58 &             -15.46 &              15.54 &    1.04\pm0.18 &      50.52\pm10.18 &     45.25\pm15.87 &  10.31 &   8.16 \\
              2.98 &             -15.62 &              15.91 &     1.17\pm0.2 &      52.22\pm10.67 &     52.81\pm15.56 &  10.33 &   8.44 \\
              4.40 &             -15.80 &              16.40 &    1.57\pm0.26 &       73.5\pm16.48 &     93.37\pm19.37 &   9.86 &   7.65 \\
              5.83 &             -15.98 &              17.01 &     1.37\pm0.3 &      65.14\pm19.49 &     75.49\pm24.09 &   7.25 &   6.33 \\
              7.29 &             -16.16 &              17.73 &    1.07\pm0.31 &      51.16\pm21.14 &      44.8\pm31.12 &   4.48 &   5.05 \\
            -28.70 &             -12.48 &              31.30 &    0.37\pm0.23 &        0.43\pm43.9 &       0.0\pm85.72 &    -- &   5.82 \\
            -27.49 &             -12.38 &              30.15 &    0.37\pm0.16 &       6.66\pm30.52 &       0.0\pm68.81 &   1.58 &   8.14 \\
            -26.27 &             -12.30 &              29.01 &    0.45\pm0.32 &    184.63\pm145.61 &   189.79\pm178.76 &   2.02 &  10.43 \\
            -25.06 &             -12.23 &              27.89 &    0.46\pm0.29 &    184.63\pm137.74 &   200.48\pm148.99 &   2.76 &  11.66 \\
            -23.86 &             -12.18 &              26.79 &    0.38\pm0.15 &     -76.85\pm22.13 &       0.0\pm58.87 &   4.42 &  11.48 \\
            -22.65 &             -12.14 &              25.70 &    0.54\pm0.16 &     -93.65\pm17.32 &       0.0\pm50.82 &   5.78 &  10.19 \\
            -21.45 &             -12.11 &              24.63 &     0.55\pm0.2 &    -101.55\pm19.68 &       0.0\pm54.45 &   5.35 &   8.56 \\
            -20.24 &             -12.10 &              23.58 &    0.55\pm0.22 &     -92.37\pm20.21 &       0.0\pm56.57 &   4.56 &   7.12 \\
            -19.04 &             -12.10 &              22.56 &    0.83\pm0.24 &     -54.58\pm18.72 &     38.72\pm30.66 &   4.81 &   5.86 \\
             -2.68 &             -13.21 &              13.48 &    0.92\pm0.23 &      51.43\pm15.78 &     41.45\pm24.27 &   5.79 &   5.47 \\
             -1.33 &             -13.36 &              13.43 &    1.15\pm0.22 &      52.58\pm14.36 &     51.15\pm20.43 &   7.34 &   6.13 \\
              0.03 &             -13.52 &              13.52 &     1.22\pm0.2 &       49.78\pm13.5 &     60.89\pm18.09 &   8.27 &   6.97 \\
              1.41 &             -13.69 &              13.76 &      1.2\pm0.2 &      56.98\pm13.41 &     72.22\pm17.07 &   8.91 &   7.60 \\
              2.81 &             -13.86 &              14.14 &    1.66\pm0.23 &       89.4\pm16.06 &    111.07\pm18.35 &  10.50 &   8.23 \\
              4.22 &             -14.03 &              14.65 &    1.79\pm0.25 &      83.64\pm17.13 &    111.91\pm19.51 &  10.45 &   7.80 \\
              5.65 &             -14.21 &              15.30 &    1.42\pm0.24 &       49.2\pm15.37 &     63.78\pm20.16 &   8.09 &   6.84 \\
              7.10 &             -14.40 &              16.05 &    1.11\pm0.24 &      40.86\pm16.28 &     33.89\pm27.06 &   5.25 &   5.88 \\
              8.57 &             -14.58 &              16.92 &    0.68\pm0.26 &      61.54\pm25.35 &      18.03\pm46.9 &   2.40 &   5.01 \\
            -28.60 &             -10.69 &              30.53 &    0.39\pm0.74 &   -186.18\pm276.28 &   120.71\pm253.41 &    -- &   5.67 \\
            -27.40 &             -10.59 &              29.37 &    0.37\pm3.58 &  -186.02\pm2740.28 &   257.12\pm1630.5 &    -- &   7.45 \\
            -26.20 &             -10.51 &              28.23 &    0.36\pm0.32 &    184.63\pm147.98 &   153.93\pm193.14 &   0.83 &   8.98 \\
            -25.00 &             -10.44 &              27.09 &    0.85\pm1.05 &  -186.18\pm1003.32 &   365.85\pm352.21 &   2.85 &   9.54 \\
            -23.81 &             -10.39 &              25.98 &    0.37\pm0.17 &    -102.85\pm24.35 &       0.0\pm61.02 &   4.35 &   9.05 \\
            -22.62 &             -10.35 &              24.87 &    0.54\pm0.18 &     -111.81\pm20.5 &       0.0\pm53.71 &   5.17 &   7.91 \\
            -21.43 &             -10.32 &              23.78 &     0.6\pm0.23 &    -108.09\pm21.84 &       0.0\pm56.63 &   4.52 &   6.36 \\
            -20.24 &             -10.31 &              22.71 &     0.6\pm0.26 &    -107.65\pm25.76 &       0.0\pm62.23 &   3.18 &   5.12 \\
             -0.14 &             -11.76 &              11.76 &     0.9\pm0.21 &       13.35\pm14.7 &       0.0\pm45.25 &   4.53 &   5.30 \\
              1.24 &             -11.93 &              11.99 &     1.45\pm0.3 &      85.75\pm28.01 &    127.51\pm31.65 &   5.99 &   5.89 \\
              2.63 &             -12.10 &              12.39 &    1.63\pm0.28 &      96.58\pm22.12 &    128.98\pm25.18 &   7.70 &   6.40 \\
              4.04 &             -12.28 &              12.93 &    1.15\pm0.23 &      59.32\pm14.48 &     63.75\pm19.21 &   7.24 &   6.55 \\
              5.47 &             -12.46 &              13.61 &    0.89\pm0.24 &      55.54\pm16.96 &     33.19\pm28.49 &   5.27 &   6.22 \\
              6.92 &             -12.65 &              14.42 &    0.59\pm0.23 &      80.13\pm25.28 &     22.97\pm44.17 &   2.76 &   5.90 \\
              8.38 &             -12.84 &              15.33 &     0.0\pm0.21 &     165.99\pm29.04 &     46.98\pm39.96 &   2.31 &   5.28 \\
            -28.50 &              -8.91 &              29.86 &    0.12\pm1.06 &   -186.18\pm974.14 &   108.78\pm764.66 &    -- &   5.14 \\
            -27.31 &              -8.81 &              28.69 &    0.33\pm7.39 &  -186.18\pm6803.49 &  288.17\pm3672.24 &    -- &   6.12 \\
            -26.12 &              -8.73 &              27.54 &     0.6\pm0.63 &    184.63\pm202.72 &   206.69\pm281.61 &   1.62 &   6.67 \\
            -24.94 &              -8.66 &              26.40 &     1.1\pm0.67 &   -186.18\pm176.53 &    219.25\pm80.39 &   3.89 &   6.56 \\
            -23.76 &              -8.60 &              25.27 &     0.5\pm0.22 &    -106.56\pm23.07 &       0.0\pm58.59 &   3.96 &   6.15 \\
            -22.58 &              -8.56 &              24.15 &     0.5\pm0.27 &    -113.12\pm28.09 &       0.0\pm64.52 &   3.39 &   5.22 \\
              6.73 &             -10.91 &              12.82 &    0.04\pm0.24 &     154.39\pm33.89 &     46.14\pm46.05 &   2.06 &   5.23 \\
              8.19 &             -11.10 &              13.80 &     0.0\pm0.23 &     157.22\pm29.67 &     61.35\pm37.77 &   2.70 &   5.14 \\
             -2.32 &              -2.99 &               3.78 &    2.06\pm0.45 &      36.27\pm45.45 &    164.81\pm41.92 &   5.53 &   5.09 \\
             -0.99 &              -3.16 &               3.31 &    2.26\pm0.72 &      18.46\pm37.74 &     154.21\pm35.4 &   6.93 &   5.73 \\
              0.36 &              -3.34 &               3.36 &    2.17\pm0.35 &      27.62\pm30.34 &    152.51\pm28.97 &   6.66 &   5.70 \\
              1.73 &              -3.52 &               3.92 &    1.47\pm0.36 &      49.72\pm37.76 &    129.76\pm38.77 &   4.31 &   5.06 \\
             -3.79 &              -1.14 &               3.95 &    2.47\pm0.44 &       9.54\pm50.65 &    204.33\pm35.09 &   7.65 &   5.62 \\
             -2.49 &              -1.31 &               2.81 &    2.59\pm0.34 &      23.23\pm29.41 &    177.13\pm24.41 &   9.85 &   6.63 \\
             -1.16 &              -1.48 &               1.89 &     2.46\pm0.5 &       24.5\pm23.07 &    151.98\pm21.71 &  11.12 &   7.41 \\
              0.18 &              -1.66 &               1.67 &    2.28\pm0.47 &       32.6\pm22.68 &     145.31\pm22.3 &  10.74 &   7.36 \\
              1.55 &              -1.84 &               2.41 &    1.86\pm0.48 &      45.46\pm32.23 &    132.66\pm27.74 &   8.35 &   6.44 \\
              2.93 &              -2.03 &               3.57 &    1.21\pm0.39 &       33.77\pm38.0 &    109.12\pm40.44 &   4.79 &   5.25 \\
             -3.95 &               0.52 &               3.99 &    2.99\pm0.42 &      -6.39\pm47.54 &    222.81\pm30.57 &  10.32 &   6.29 \\
             -2.66 &               0.35 &               2.68 &    2.97\pm0.32 &      15.38\pm28.52 &    196.49\pm21.93 &  12.55 &   7.36 \\
             -1.34 &               0.18 &               1.35 &    2.74\pm0.31 &      18.32\pm22.46 &    168.65\pm18.79 &  13.99 &   8.17 \\
              0.00 &               0.00 &               0.00 &      2.7\pm0.3 &      15.73\pm21.07 &    165.89\pm18.14 &  14.28 &   8.20 \\
              1.36 &              -0.18 &               1.37 &    2.69\pm0.29 &       8.55\pm21.77 &    170.93\pm19.24 &  12.70 &   7.18 \\
              2.75 &              -0.37 &               2.77 &     2.2\pm0.31 &     -14.19\pm26.05 &    158.32\pm24.11 &   8.85 &   5.99 \\
              4.15 &              -0.56 &               4.19 &    1.46\pm0.41 &     -71.27\pm43.86 &    135.18\pm45.41 &   4.54 &   5.26 \\
             -4.12 &               2.17 &               4.66 &     3.1\pm0.37 &      21.45\pm38.74 &     221.9\pm28.06 &  10.28 &   5.77 \\
             -2.83 &               2.00 &               3.47 &     2.9\pm0.33 &      19.85\pm27.93 &    193.01\pm22.56 &  11.85 &   6.82 \\
             -1.52 &               1.83 &               2.37 &    2.73\pm0.31 &       8.34\pm22.81 &    174.68\pm19.65 &  13.13 &   7.60 \\
             -0.18 &               1.65 &               1.66 &    2.72\pm0.29 &       1.43\pm20.02 &     170.1\pm17.54 &  13.79 &   7.54 \\
              1.18 &               1.46 &               1.88 &     2.66\pm0.3 &       -1.86\pm19.6 &    159.09\pm18.19 &  12.94 &   6.87 \\
              2.56 &               1.28 &               2.86 &    2.14\pm0.33 &      -7.67\pm20.56 &    127.22\pm22.25 &   9.85 &   5.89 \\
              3.96 &               1.09 &               4.11 &     1.4\pm0.36 &     -34.95\pm29.57 &    110.37\pm33.62 &   5.57 &   5.18 \\
             -3.00 &               3.63 &               4.71 &     2.32\pm0.4 &      22.59\pm29.23 &    156.62\pm27.26 &   8.34 &   5.27 \\
             -1.69 &               3.46 &               3.85 &    2.11\pm0.38 &       4.18\pm27.46 &    147.42\pm26.48 &   8.92 &   6.04 \\
             -0.36 &               3.28 &               3.30 &    2.06\pm0.33 &     -13.38\pm23.62 &     150.65\pm22.6 &   9.72 &   6.13 \\
              0.99 &               3.10 &               3.25 &     2.06\pm0.4 &     -33.94\pm24.13 &    150.46\pm22.07 &   9.68 &   5.84 \\
              2.37 &               2.91 &               3.75 &    1.71\pm0.37 &      -35.0\pm25.12 &    123.73\pm27.33 &   7.40 &   5.41 \\
             -0.54 &               4.89 &               4.92 &    1.07\pm0.38 &     -27.83\pm26.82 &    109.67\pm29.41 &   6.14 &   5.14 \\
              0.81 &               4.71 &               4.78 &    0.37\pm0.45 &     -62.42\pm21.87 &     88.75\pm26.87 &   5.62 &   5.03 \\
             -4.62 &               7.02 &               8.40 &    0.93\pm0.23 &      28.83\pm13.88 &       0.0\pm43.85 &   5.05 &   5.06 \\
             -3.35 &               6.85 &               7.62 &    0.79\pm0.25 &      12.41\pm18.35 &     38.77\pm28.53 &   4.94 &   5.47 \\
             -2.05 &               6.67 &               6.98 &    0.74\pm0.26 &      -1.21\pm18.99 &     61.85\pm24.84 &   5.57 &   5.56 \\
             -0.72 &               6.49 &               6.53 &    0.87\pm0.28 &     -15.61\pm19.73 &     82.75\pm23.54 &   6.14 &   5.56 \\
              0.63 &               6.31 &               6.35 &    0.01\pm0.27 &     -52.08\pm17.93 &     47.25\pm25.97 &   4.62 &   5.22 \\
             -7.26 &               8.92 &              11.50 &     0.9\pm0.21 &      10.17\pm12.39 &     26.95\pm23.65 &   6.38 &   5.81 \\
             -6.04 &               8.76 &              10.64 &    1.07\pm0.19 &       20.65\pm9.47 &      14.9\pm24.31 &   8.43 &   7.12 \\
             -4.79 &               8.59 &               9.84 &     1.0\pm0.17 &       16.79\pm8.94 &        0.0\pm34.7 &   9.51 &   8.28 \\
             -3.52 &               8.42 &               9.13 &    0.93\pm0.16 &        7.57\pm9.45 &      26.8\pm18.89 &  10.65 &   8.95 \\
             -2.22 &               8.25 &               8.55 &    1.04\pm0.17 &        0.93\pm9.85 &     47.86\pm14.86 &  11.64 &   8.85 \\
             -0.90 &               8.08 &               8.13 &     1.2\pm0.18 &       3.81\pm11.58 &     74.61\pm14.57 &  10.91 &   8.41 \\
              0.45 &               7.90 &               7.91 &    1.45\pm0.27 &      -4.45\pm21.11 &     126.09\pm21.5 &   8.93 &   7.42 \\
              1.82 &               7.72 &               7.93 &    1.63\pm0.42 &     -56.86\pm41.78 &    170.17\pm30.83 &   6.92 &   6.19 \\
             -8.61 &              10.64 &              13.69 &    0.82\pm0.19 &      13.02\pm14.17 &     30.92\pm24.97 &   5.29 &   6.34 \\
             -7.42 &              10.48 &              12.84 &    1.06\pm0.14 &       14.69\pm8.11 &     34.56\pm14.57 &  10.22 &   9.03 \\
             -6.20 &              10.32 &              12.04 &    1.06\pm0.13 &       20.37\pm6.89 &      25.2\pm14.95 &  13.71 &  11.85 \\
             -4.96 &              10.15 &              11.30 &    1.05\pm0.12 &       18.59\pm6.62 &     21.74\pm15.58 &  17.57 &  14.18 \\
             -3.69 &               9.98 &              10.65 &    1.13\pm0.12 &       15.16\pm6.54 &      35.4\pm11.84 &  21.12 &  15.78 \\
             -2.40 &               9.81 &              10.10 &    1.25\pm0.12 &       12.72\pm6.74 &     49.54\pm10.22 &  22.46 &  16.20 \\
             -1.08 &               9.64 &               9.70 &    1.31\pm0.14 &       14.89\pm7.98 &      64.47\pm10.7 &  20.40 &  14.82 \\
              0.27 &               9.47 &               9.47 &    1.31\pm0.17 &      12.11\pm11.27 &     86.11\pm13.52 &  15.95 &  12.27 \\
              1.64 &               9.29 &               9.43 &    1.43\pm0.25 &     -17.79\pm18.61 &    118.96\pm19.65 &  11.89 &   9.53 \\
              3.03 &               9.11 &               9.60 &    1.39\pm0.36 &     -34.08\pm26.56 &    125.69\pm26.57 &   8.47 &   7.37 \\
              4.44 &               8.93 &               9.97 &     0.0\pm0.58 &     -33.66\pm38.47 &    102.27\pm43.36 &   3.58 &   5.31 \\
             -9.93 &              12.33 &              15.83 &    0.76\pm0.24 &      64.43\pm22.46 &      15.2\pm44.67 &   2.87 &   5.36 \\
             -8.77 &              12.17 &              15.00 &    1.05\pm0.16 &      21.52\pm10.83 &      37.1\pm18.24 &   8.26 &   8.62 \\
             -7.58 &              12.02 &              14.21 &     1.1\pm0.12 &       15.08\pm7.76 &     46.23\pm12.09 &  14.33 &  13.00 \\
             -6.37 &              11.85 &              13.46 &    1.08\pm0.11 &        16.6\pm6.66 &     44.83\pm10.63 &  20.30 &  17.74 \\
             -5.13 &              11.69 &              12.77 &    1.11\pm0.11 &       18.66\pm6.03 &     39.09\pm10.41 &  27.15 &  21.99 \\
             -3.86 &              11.52 &              12.15 &     1.18\pm0.1 &        18.59\pm5.7 &      40.72\pm9.66 &  33.15 &  25.27 \\
             -2.56 &              11.36 &              11.64 &     1.27\pm0.1 &       15.68\pm5.79 &      51.46\pm8.67 &  35.87 &  26.28 \\
             -1.25 &              11.19 &              11.26 &    1.34\pm0.12 &       13.85\pm6.73 &      65.43\pm9.01 &  34.00 &  24.21 \\
              0.10 &              11.02 &              11.02 &    1.38\pm0.14 &        11.88\pm8.2 &     76.88\pm10.29 &  28.14 &  19.87 \\
              1.46 &              10.85 &              10.95 &    1.48\pm0.17 &       2.11\pm10.37 &     91.57\pm12.25 &  21.62 &  15.11 \\
              2.85 &              10.67 &              11.05 &    1.52\pm0.22 &      -8.68\pm13.19 &    101.82\pm14.95 &  15.80 &  10.90 \\
              4.26 &              10.49 &              11.32 &    1.33\pm0.28 &      -4.13\pm17.86 &    103.21\pm19.91 &   9.85 &   7.65 \\
              5.70 &              10.31 &              11.78 &     0.87\pm0.4 &       9.88\pm35.83 &     105.7\pm39.01 &   4.25 &   5.22 \\
            -10.08 &              13.84 &              17.12 &     0.8\pm0.21 &      42.71\pm17.99 &       0.0\pm50.39 &   3.82 &   5.97 \\
             -8.92 &              13.69 &              16.34 &    1.04\pm0.15 &       12.9\pm10.09 &     39.49\pm16.63 &   9.85 &  10.11 \\
             -7.74 &              13.53 &              15.59 &     1.1\pm0.12 &       10.51\pm8.27 &     52.87\pm12.05 &  17.37 &  16.00 \\
             -6.53 &              13.37 &              14.88 &    1.09\pm0.11 &       10.94\pm6.81 &     49.15\pm10.36 &  25.74 &  22.22 \\
             -5.29 &              13.21 &              14.23 &     1.13\pm0.1 &       13.57\pm5.73 &       43.8\pm9.32 &  34.72 &  28.31 \\
             -4.02 &              13.04 &              13.65 &      1.2\pm0.1 &       13.93\pm5.42 &      45.91\pm8.63 &  43.04 &  33.18 \\
             -2.73 &              12.88 &              13.17 &     1.28\pm0.1 &       11.66\pm5.67 &      55.92\pm8.17 &  48.28 &  35.81 \\
             -1.41 &              12.72 &              12.79 &    1.36\pm0.11 &        9.66\pm6.19 &      67.75\pm8.19 &  48.11 &  33.94 \\
             -0.07 &              12.55 &              12.55 &    1.45\pm0.12 &        8.13\pm6.92 &      76.22\pm8.75 &  42.15 &  28.66 \\
              1.29 &              12.39 &              12.45 &    1.54\pm0.14 &        4.62\pm7.84 &      83.56\pm9.58 &  33.25 &  22.05 \\
              2.67 &              12.22 &              12.51 &    1.55\pm0.15 &       -1.29\pm8.73 &     88.71\pm10.47 &  24.17 &  15.88 \\
              4.08 &              12.04 &              12.71 &    1.37\pm0.17 &        1.85\pm9.63 &     82.54\pm11.77 &  15.74 &  10.82 \\
              5.51 &              11.86 &              13.08 &    1.07\pm0.22 &      12.55\pm13.75 &     72.49\pm17.31 &   8.84 &   7.30 \\
            -10.23 &              15.34 &              18.43 &    0.63\pm0.21 &      -7.73\pm20.77 &       0.0\pm54.64 &   3.16 &   5.68 \\
             -9.08 &              15.18 &              17.69 &    1.01\pm0.15 &      -3.23\pm10.25 &     37.25\pm17.31 &   9.82 &   9.87 \\
             -7.90 &              15.02 &              16.97 &    1.12\pm0.12 &        5.69\pm7.87 &     50.63\pm11.73 &  17.89 &  16.00 \\
             -6.69 &              14.86 &              16.30 &     1.07\pm0.1 &        6.94\pm6.53 &     47.59\pm10.12 &  26.61 &  22.92 \\
             -5.45 &              14.70 &              15.68 &     1.1\pm0.09 &        8.05\pm5.66 &      45.55\pm9.02 &  35.99 &  29.49 \\
             -4.19 &              14.54 &              15.13 &    1.18\pm0.09 &        8.33\pm5.43 &      48.23\pm8.43 &  45.47 &  35.58 \\
             -2.89 &              14.38 &              14.67 &     1.26\pm0.1 &        8.46\pm5.74 &      56.56\pm8.22 &  52.88 &  39.36 \\
             -1.57 &              14.22 &              14.31 &    1.37\pm0.11 &        7.95\pm6.06 &       66.1\pm8.11 &  55.52 &  39.30 \\
             -0.23 &              14.07 &              14.07 &    1.48\pm0.11 &         5.96\pm6.4 &       72.8\pm8.24 &  51.37 &  35.12 \\
              1.13 &              13.91 &              13.95 &    1.53\pm0.12 &        3.05\pm6.66 &      76.56\pm8.42 &  41.59 &  28.01 \\
              2.51 &              13.75 &              13.97 &    1.53\pm0.12 &        1.32\pm6.84 &      76.84\pm8.64 &  30.52 &  20.58 \\
              3.90 &              13.58 &              14.13 &    1.42\pm0.13 &         2.8\pm7.42 &      73.68\pm9.49 &  20.61 &  14.20 \\
              5.33 &              13.40 &              14.42 &    1.24\pm0.16 &         9.27\pm9.1 &     72.26\pm11.63 &  13.03 &   9.44 \\
              6.78 &              13.21 &              14.85 &    1.14\pm0.24 &      20.12\pm14.12 &     80.54\pm17.14 &   7.88 &   6.03 \\
             -9.24 &              16.65 &              19.04 &    1.08\pm0.17 &      -6.97\pm12.29 &     47.19\pm18.37 &   8.44 &   8.13 \\
             -8.06 &              16.49 &              18.35 &    1.17\pm0.13 &        5.27\pm8.51 &     52.77\pm12.39 &  15.31 &  13.12 \\
             -6.85 &              16.33 &              17.71 &    1.04\pm0.11 &        4.34\pm6.99 &     49.17\pm10.62 &  21.76 &  18.88 \\
             -5.61 &              16.17 &              17.11 &    1.04\pm0.09 &        4.25\pm6.05 &      45.84\pm9.58 &  29.02 &  25.05 \\
             -4.34 &              16.01 &              16.59 &    1.12\pm0.09 &        6.71\pm5.53 &      45.17\pm8.87 &  37.76 &  30.84 \\
             -3.05 &              15.85 &              16.14 &     1.22\pm0.1 &        8.76\pm5.57 &      50.74\pm8.42 &  46.35 &  35.46 \\
             -1.72 &              15.70 &              15.79 &     1.34\pm0.1 &        7.06\pm5.73 &       58.39\pm8.1 &  51.99 &  37.62 \\
             -0.38 &              15.55 &              15.56 &    1.42\pm0.11 &        2.97\pm5.92 &      64.46\pm8.02 &  51.12 &  35.94 \\
              0.98 &              15.41 &              15.44 &    1.46\pm0.11 &         0.8\pm5.95 &      67.79\pm7.89 &  43.95 &  30.57 \\
              2.35 &              15.26 &              15.44 &    1.48\pm0.11 &        0.75\pm5.96 &      68.94\pm7.86 &  33.91 &  23.54 \\
              3.73 &              15.10 &              15.55 &    1.41\pm0.11 &        0.56\pm6.54 &      69.77\pm8.56 &  23.45 &  16.50 \\
              5.14 &              14.92 &              15.78 &    1.28\pm0.14 &        6.75\pm7.83 &     70.51\pm10.13 &  15.17 &  10.92 \\
              6.59 &              14.72 &              16.13 &    1.21\pm0.19 &      21.41\pm11.29 &     70.78\pm14.42 &   9.33 &   7.02 \\
             -9.39 &              18.09 &              20.38 &    1.03\pm0.25 &      -9.87\pm16.49 &     41.68\pm25.21 &   6.03 &   5.46 \\
             -8.22 &              17.93 &              19.73 &    1.07\pm0.18 &       -1.6\pm11.92 &     48.69\pm17.62 &  10.17 &   8.60 \\
             -7.02 &              17.77 &              19.10 &    0.89\pm0.13 &       -1.24\pm9.54 &     45.67\pm14.73 &  13.22 &  12.61 \\
             -5.77 &              17.61 &              18.53 &      0.9\pm0.1 &        0.82\pm7.19 &     37.89\pm12.44 &  17.64 &  17.08 \\
             -4.50 &              17.45 &              18.02 &    1.03\pm0.09 &        6.67\pm5.77 &     36.56\pm10.35 &  25.09 &  21.86 \\
             -3.20 &              17.29 &              17.59 &    1.16\pm0.09 &        9.68\pm5.42 &      41.41\pm9.12 &  33.54 &  26.56 \\
             -1.87 &              17.15 &              17.25 &     1.27\pm0.1 &         6.2\pm5.44 &      47.57\pm8.51 &  40.25 &  30.15 \\
             -0.52 &              17.01 &              17.02 &     1.34\pm0.1 &        0.49\pm5.47 &      55.42\pm7.93 &  43.06 &  31.31 \\
              0.84 &              16.88 &              16.90 &     1.39\pm0.1 &       -1.29\pm5.43 &      60.96\pm7.54 &  41.08 &  28.89 \\
              2.20 &              16.76 &              16.90 &      1.4\pm0.1 &       -1.23\pm5.39 &      60.85\pm7.49 &  34.11 &  23.45 \\
              3.55 &              16.61 &              16.98 &    1.34\pm0.11 &       -1.01\pm5.87 &      58.92\pm8.26 &  24.30 &  17.08 \\
              4.94 &              16.41 &              17.14 &    1.23\pm0.14 &         4.36\pm7.7 &     60.09\pm10.65 &  15.15 &  11.42 \\
              6.39 &              16.20 &              17.42 &    1.09\pm0.19 &      16.02\pm12.13 &     60.18\pm16.41 &   8.09 &   7.16 \\
             -7.18 &              19.19 &              20.48 &    0.64\pm0.19 &     -15.47\pm15.25 &     26.98\pm27.96 &   6.26 &   7.04 \\
             -5.94 &              19.02 &              19.93 &    0.73\pm0.14 &      -7.14\pm10.23 &     18.22\pm23.65 &   8.97 &   9.89 \\
             -4.66 &              18.86 &              19.43 &    0.92\pm0.11 &        2.88\pm6.73 &     22.86\pm15.28 &  14.56 &  13.24 \\
             -3.35 &              18.70 &              19.00 &     1.1\pm0.11 &         7.14\pm6.1 &     33.01\pm11.53 &  21.01 &  17.18 \\
             -2.01 &              18.55 &              18.66 &     1.21\pm0.1 &        3.78\pm5.89 &      40.66\pm9.97 &  26.95 &  21.35 \\
             -0.65 &              18.42 &              18.43 &     1.29\pm0.1 &       -2.28\pm5.37 &      50.67\pm8.15 &  32.17 &  24.13 \\
              0.73 &              18.30 &              18.32 &     1.33\pm0.1 &       -4.87\pm5.13 &      55.22\pm7.47 &  34.09 &  24.01 \\
              2.09 &              18.22 &              18.34 &     1.29\pm0.1 &       -3.66\pm5.09 &       49.4\pm7.84 &  30.03 &  20.77 \\
              3.37 &              18.10 &              18.41 &    1.21\pm0.11 &        -1.2\pm5.53 &      41.76\pm9.29 &  21.89 &  15.61 \\
              4.71 &              17.86 &              18.47 &    1.15\pm0.15 &         0.9\pm7.81 &      42.9\pm12.68 &  13.27 &  10.54 \\
              6.17 &              17.63 &              18.68 &    0.91\pm0.21 &       1.42\pm15.36 &      44.07\pm23.2 &   5.92 &   6.50 \\
             -6.10 &              20.41 &              21.30 &    0.71\pm0.24 &     -12.52\pm18.58 &     29.37\pm31.88 &   4.39 &   5.24 \\
             -4.82 &              20.25 &              20.81 &    0.84\pm0.16 &       -0.98\pm10.3 &     20.66\pm22.64 &   7.98 &   7.44 \\
             -3.51 &              20.08 &              20.39 &    1.02\pm0.13 &       -2.73\pm7.83 &     24.41\pm16.81 &  12.04 &  10.64 \\
             -2.16 &              19.92 &              20.04 &    1.16\pm0.11 &       -6.59\pm6.55 &     34.32\pm12.06 &  16.93 &  14.25 \\
             -0.79 &              19.76 &              19.78 &     1.24\pm0.1 &      -10.46\pm5.74 &      43.69\pm9.37 &  22.00 &  17.31 \\
              0.62 &              19.61 &              19.62 &     1.24\pm0.1 &      -10.93\pm5.77 &      47.87\pm8.98 &  24.23 &  18.27 \\
              2.05 &              19.48 &              19.59 &    1.17\pm0.11 &       -6.19\pm6.22 &     43.38\pm10.15 &  21.74 &  16.45 \\
              3.02 &              19.42 &              19.65 &    1.11\pm0.13 &        0.25\pm6.74 &     33.76\pm12.55 &  16.22 &  12.96 \\
              4.47 &              19.13 &              19.65 &    1.14\pm0.17 &        3.47\pm9.36 &      35.25\pm16.5 &  10.24 &   8.83 \\
              5.99 &              18.97 &              19.89 &    0.99\pm0.25 &      -3.41\pm18.06 &      38.75\pm28.3 &   4.52 &   5.52 \\
             -3.68 &              21.44 &              21.75 &    0.84\pm0.19 &      -19.9\pm12.44 &       0.0\pm41.13 &   6.07 &   6.44 \\
             -2.33 &              21.27 &              21.39 &    1.03\pm0.14 &      -16.25\pm7.86 &       8.07\pm25.2 &  10.01 &   9.20 \\
             -0.96 &              21.08 &              21.10 &     1.1\pm0.13 &       -13.5\pm7.23 &     26.17\pm15.19 &  13.68 &  11.82 \\
              0.43 &              20.88 &              20.88 &    1.07\pm0.13 &      -12.43\pm8.24 &     40.84\pm13.59 &  14.73 &  12.94 \\
              1.79 &              20.63 &              20.71 &    1.04\pm0.14 &      -7.69\pm10.08 &     54.91\pm14.34 &  12.85 &  12.31 \\
              3.04 &              20.43 &              20.65 &    1.06\pm0.15 &         5.35\pm9.8 &     42.89\pm15.59 &   9.92 &   9.87 \\
              4.39 &              20.38 &              20.84 &    1.22\pm0.19 &      10.09\pm12.05 &     38.32\pm19.77 &   7.03 &   6.91 \\
             -2.52 &              22.61 &              22.75 &    0.74\pm0.22 &     -13.84\pm17.95 &       0.0\pm50.39 &   4.02 &   5.75 \\
             -1.16 &              22.42 &              22.45 &    0.89\pm0.18 &       -1.3\pm12.27 &        0.0\pm40.9 &   6.70 &   7.75 \\
              0.21 &              22.23 &              22.23 &    0.84\pm0.17 &      -1.64\pm14.06 &     36.13\pm23.24 &   7.27 &   8.78 \\
              1.57 &              22.03 &              22.09 &     0.82\pm0.2 &      -0.63\pm20.76 &     67.59\pm26.42 &   5.85 &   8.35 \\
              2.92 &              21.88 &              22.07 &    0.88\pm0.19 &      11.57\pm16.88 &     37.77\pm26.74 &   4.52 &   6.80 \\
              0.03 &              23.62 &              23.62 &     0.6\pm0.25 &       1.75\pm28.58 &      6.44\pm60.09 &   2.00 &   5.22 \\
              1.40 &              23.46 &              23.50 &    0.47\pm0.27 &     -11.35\pm40.72 &       0.0\pm81.57 &    -- &   5.02 \\
\enddata
\tablenotetext{}{$^a$Arc-position physical separation to galaxy in the absorber plane; $^b$Projected physical separation to galaxy in the absorber plane; $^c$\MgII $\lambda$2796 absorption-strength (with 1$\sigma$ error); $^d$Velocity offset relative to $z=0.7711$ (with 1$\sigma$ error); $^e$Velocity dispersion relative to $z=0.7711$ (with 1$\sigma$ error); $^f$\textcolor{black}{\MgII absorption significance values}; $^g$Signal-to-noise ratio to the continuum.}
\tablenotetext{}{\textcolor{black}{NOTE: Non-physical \MgII absorption significance values are left blank.}}
\end{deluxetable*}



\end{document}